\documentclass[aps,groupedaddress
,amsfonts,amsmath,floatfix,eqsecnum,nofootinbib
]{revtex4}

\usepackage[usenames]{color}

\usepackage{epsfig}
\usepackage{graphicx}
\usepackage{mathptmx}      % use Times fonts if available on your TeX system
\usepackage{latexsym}
\usepackage{bm}
\usepackage{amstext}\usepackage{array}

\def\slashchar#1{\setbox0=\hbox{$#1$}
   \dimen0=\wd0 \setbox1=\hbox{/} \dimen1=\wd1
   \ifdim\dimen0>\dimen1 \rlap{\hbox to \dimen0{\hfil/\hfil}} #1
   \else  \rlap{\hbox to \dimen1{\hfil$#1$\hfil}} / \fi}

\newcommand{\cO}{{\mathcal{O}}}

\newcommand{\U}{{\mathrm{U}}}
\newcommand{\Eq}[1]{Eq.~(\ref{eq:#1})}

\newcommand{\vv}{{\bm{v}}}

\newcommand{\vnabla}{{\bm{\nabla}}}
\newcommand{\ignore}[1]{}

\newcommand{\R}{\mathbb{R}}
\newcommand{\C}{\mathbb{C}}
\newcommand{\Z}{\mathbb{Z}}
\renewcommand{\Re}{{\mathrm{Re}}\,}
\renewcommand{\Im}{{\mathrm{Im}}\,}

\newcommand{\esp}[1]{{\langle #1 \rangle}}

\newcommand{\A}{{\mathcal{A}}}
\newcommand{\tA}{{\tilde{\A}}}

\begin{document}

\title{Does the complex Langevin method give unbiased results?}

\author{L.L. Salcedo}
\email{salcedo@ugr.es}

\affiliation{Departamento de F\'{\i}sica At\'omica, Molecular y Nuclear and \\
  Instituto Carlos I de F\'{\i}sica Te\'orica y Computacional, \\ Universidad
  de Granada, E-18071 Granada, Spain.}

\date{\today}

\begin{abstract}
We investigate whether the stationary solution of the Fokker-Planck
equation of the complex Langevin algorithm reproduces the correct expectation
values. When the complex Langevin algorithm for an action $S(x)$ is
convergent, it produces an equivalent complex probability distribution $P(x)$
which ideally would coincide with $e^{-S(x)}$. We show that the projected
Fokker-Planck equation fulfilled by $P(x)$ may contain an anomalous term whose
form is made explicit. Such term spoils the relation $P(x)=e^{-S(x)}$,
introducing a bias in the expectation values. Through the analysis of several
periodic and non-periodic one-dimensional problems, using either exact or
numerical solutions of the Fokker-Planck equation on the complex plane, it is
shown that the anomaly is present quite generally. In fact, an anomaly is
expected whenever the Langevin walker needs only a finite time to go to
infinity and come back, and this is the case for typical actions. We
conjecture that the anomaly is the rule rather than the exception in the
one-dimensional case, however, this could change as the number of variables
involved increases.
\end{abstract}

%\pacs{ 05.10.Ln, 02.70.-c, 02.70.Ss, 12.38.Gc }
%\keywords{}

\maketitle

\tableofcontents

\section{Introduction}
\label{sec:intro}

As it is well known, in the functional integral formulation of quantum field
theories the problem is formally transformed into one of classical statistical
mechanics where the action plays the role of Hamiltonian in one more spatial
dimension. This entails to Wick rotate to Euclidean time, where the functional
integral has a better mathematical behavior \cite{Glimm:1987ng}.  In many
cases this is sufficient to make the equivalent Boltzmann weight of the
integral real and positive, thereby allowing to apply Monte Carlo techniques
in numerical calculations.  Unfortunately, even in the Euclidean formulation a
positive weight is not always guaranteed and some times one has to deal with
complex weights. A typical example comes from the introduction of a baryon
number chemical potential in lattice QCD \cite{Hasenfratz:1983ba}.  Whether in
field theory or in statistical mechanics, real but non positive or more
generally complex weights are present in some problems. This is the famous
sign (or phase) problem. Although the partition function is well defined, a
straight application of the Monte Carlo method is not available, and this
makes the problem hard to attack numerically \cite{Troyer:2004ge}.  Many ideas
have been proposed to treat the sign problem (see e.g.
\cite{Parisi:1984cs,Hamber:1985qh,Bhanot:1987nv,Salcedo:1993tj%
  ,Kieu:1993gw,Salcedo:1996sa,Barbour:1997ej,Chandrasekharan:1999cm%
  ,Fodor:2001au,Prokofiev:2001,Baeurle:2002,Moreira:2003,Azcoiti:2004ri%
  ,Berges:2005yt,Aarts:2008wh,Banerjee:2010kc,Bloch:2011jx%
  ,Cristoforetti:2013wha,Rota:2015,Salcedo:2015jxd}), with success in
particular cases, but it is fair to say that no efficient, systematic, and
robust approach exists yet to deal with this problem.

A very general and mathematically sound approach exists to deal with complex
weights, this is the reweighting technique \cite{Liu:2001}. Unfortunately, the
method suffers from the ubiquitous overlap problem: by using samples from a
different weight, importance sampling is violated and the problem worsens for
large systems, precisely where the Monte Carlo approach would be the most
efficient (or less inefficient) method. Another more or less general method is
the complex Langevin approach of Parisi
\cite{Parisi:1984cs}. Unlike reweighting, this method only
applies to continuous degrees of freedom, and moreover the action of the
system must admit a holomorphic extension to the complexified version of the
real manifold of physical configurations. Nevertheless, analyticity is not a
stringent restriction for typical actions. A further recent technique is
based on integration on Lefschetz thimbles \cite{Cristoforetti:2012su}. We
comment on this in Sec.~\ref{sec:con}.

The complex Langevin approach shares an important property with reweighting,
namely, it does not spoil the locality properties of the system to be
simulated. By locality of an action $S(x)$ [$x=(x^1,\ldots,x^n)$ is the
  configuration] we mean that for any variable $x^i$, $S(x)$ can be written as
$S_1(x)+S_2(x)$ where $S_1$ does not depend on $x^i$ and $S_2$ depends on
$x^i$ and a small number of other variables (the so-called neighbors of
$x^i$). This requirement is often crucial for an efficient implementation of a
Monte Carlo algorithm. The complex Langevin algorithm is indeed a very
intuitive and handy approach which from the beginning attracted much attention
\cite{Huffel:1984mq,Barbour:1986jf,Ambjorn:1986fz,Damgaard:1987rr%
  ,Berges:2006xc,Fukushima:2010bq}. Regrettably, unlike the real Langevin
method, its complex version has no sound mathematical foundation, and many
examples have been found where the method does not converge or converges to a
wrong equilibrium solution \cite{Ambjorn:1985iw%
  ,Ambjorn:1985cv,Salcedo:1993tj,Aarts:2011ax,Pehlevan:2007eq,Bloch:2015coa,Makino:2015ooa}.

In a typical application with a complex action $S(x)$, one needs to compute
expectation values of the complex distribution $P(x)= e^{-S(x)}$ defined on
the manifold of configurations of the physical system.  In the complex
Langevin algorithm walkers move on the complexified manifold producing some
normalized probability density $\rho(z,t)$ there. The whole point of the
approach is that after equilibrium has been reached, the stationary density
$\rho(z)$ should reproduce the correct expectation values of $P(x)$, in the
sense of analytical extension of the observables. That is,
\begin{equation}
\esp{A(x)}_P = \esp{A(z)}_\rho
,
\label{eq:2.100}
\end{equation}
where $A(z)$ refers to holomorphic extension of the observable $A(x)$, and
$\esp{A}_P$ and $\esp{A}_\rho$ stand for the expectation values on the real
manifold with $P$, and on the complexified manifold with $\rho$, respectively.

Following a suggestion in \cite{Ambjorn:1985cv}, S\"oderberg first pointed out
\cite{Soderberg:1987pd} that other densities $\rho(z)$ could exist fulfilling
the requirement (\ref{eq:2.100}) not necessarily derived from a complex
Langevin approach, and also properties of them were studied there. In
\cite{Salcedo:1996sa} such valid densities $\rho(z)$ defined on the
complexified manifold were named {\em representations} (of the target complex
probability $P(x)$). In that work properties of the representations were
analyzed and explicit constructions were carried out for $P(x)$ of the form
Gaussian times polynomial of any degree and in any number of dimensions, among
other. Necessary and sufficient conditions for the existence of
representations were established in \cite{Weingarten:2002xs}.  That
representations exist quite generally, not only on $\R^n$ but also on Lie
groups was shown in \cite{Salcedo:2007ji}, with some explicit
constructions. Representations beyond complex Langevin have recently been
considered in \cite{Wosiek:2015iwl,Wosiek:2015bqg}. Even more recently, a
Gibbs sampling approach to complex probabilities using representations has
been described in \cite{Salcedo:2015jxd}.

There are complex actions for which the complex Langevin algorithm fails to
provide a proper representation density $\rho(z)$. The above considerations
show that this is not an intrinsic problem of those actions, but rather a
limitation of the complex Langevin approach. One such example is the
one-dimensional action $S(x) = i\beta_I \cos(x)$ with $\beta_I$ real and
different from zero.  Complex Langevin arrives in this case to a $\rho(z)$ on
the complex plane which is independent of $x$, thereby predicting vanishing
expectation values for all $e^{ikx}$ except $k=0$. Such incorrect result is
not due to a pathology of this action. Indeed, as shown in
\cite{Salcedo:2015jxd}, just any periodic one-dimensional action can be
represented on the complex plane by a positive probability density of the form
\begin{equation}
\rho(z)  = Q_1(x)\delta(y-Y) + Q_2(x)\delta(y+Y)
,
\label{eq:2.102}
\end{equation}
where the periodic functions $Q_{1,2}(x)$ are positive for real $x$. By
decomposing in Fourier modes it can be easily shown that the condition
(\ref{eq:2.100}) is fulfilled by taking
\begin{equation}
Q_{1,k} = \frac{
e^{ k Y} P_k - e^{ -k Y} P^*_{-k} }{2 \sinh(2 k Y )} 
,\qquad
Q_{2,k} = - \frac{
e^{- k Y} P_k - e^{k Y} P^*_{-k} }{2 \sinh(2kY)} 
,\qquad
k\not = 0
.
\label{eq:2.101}
\end{equation}
Clearly these modes correspond to real functions $Q_{1,2}(x)$. Moreover,
adding suitable zero modes, such $Q_{1,2}(x)$ are positive by taking $Y$ above
some critical value which depends on the concrete complex action. The same
formulas (\ref{eq:2.101}) apply to non-periodic actions too, and can be
extended to any number of dimensions.  We refer to \cite{Salcedo:2015jxd} for
explicit constructions.

The construction in \Eq{2.102} is a representation with support on two lines
parallel to the real axis, and we have noted that positivity of $\rho$
requires the separation $2Y$ to be larger than some critical value dependent
on $P(x)$. This result is quite general: as a rule, the more complex a target
complex probability $P(x)$ is, the more spread on the complex plane should be
any representation of it. For instance, if a {\em real} observable develops a
complex expectation value (due to the complex action), such expectation value
will not be reproduced by a density $\rho(z)$ lying too close to the real axis
(where the observable is real). This observation can be made more precise: For
any observable $A(x)$, let $\mathbf{A}$ be the set of points $z$ in the
complexified manifold such that $|A(z)| \ge |\esp{A}_P|$. Then clearly the
support of any valid representation $\rho$ of $P$ should have some overlap
with $\mathbf{A}$, since otherwise $|\esp{A}_\rho|< |\esp{A}_P|$.

The overlap condition just mentioned can be applied immediately to the the
complex Langevin discussion, through an example noted in
\cite{Salcedo:2015jxd}. Consider the one-dimensional action $S(x) =
(\beta/4)x^4+iqx$, with positive $\beta$ and $q$. In one applies a standard
complex Langevin approach here, the walkers will be subject to the usual
horizontal diffusion plus a drift with horizontal and vertical components.
However, on the real axis the vertical drift is purely downwards, because the
term with $\beta$ does not contribute there. This implies that at equilibrium
all the walkers will end up in the lower half plane, since once they move
there they have no way to cross the real axis again. It follows that the
support of the equilibrium complex Langevin process, $\rho_{\rm CL}(z)$ is
entirely contained in $\{y\le 0\}$. On the other hand, for $k>0$, $|e^{-ikz}|
= e^{ky}\le 1$ if $y\le 0$, therefore $|\esp{e^{-ikz}}_{\rm CL}| \le 1$. Yet
for $\beta=1/2$, $q=2$, $\esp{e^{-ix}}=-4.98$. So we can conclude that, for
this action, the complex Langevin algorithm is necessarily converging to the
wrong equilibrium distribution.

It is also known that the Langevin algorithm is afflicted by the segregation
problem \cite{Fujimura:1993cq}, so it cannot be used to represent a real but
non positive weight such as $P(x) = 1+2\cos(x)$. We emphasize that there is
nothing intrinsically wrong with this $P(x)$ or with the $(\beta/4)x^4+iqx$
action above. The two cases can be represented, using for instance the
two-branch approach of \Eq{2.102}.

The complex Langevin method has captivated many a researcher due to its
elegance \cite{Gausterer:1998jw,Adami:2000fs}.  Even some tentative proofs of
its validity were advanced (see e.g. \cite{Gozzi:1984qv,Lee:1994swa}).
Besides, the method has achieved some empirical success in concrete problems
\cite{Gausterer:1985sm,Aarts:2008wh}. Nevertheless, beautiful does not imply
correct and after 30 years the mathematical basis of the method has not been
established beyond the rather trivial cases of quadratic actions or real
actions.  Convergence requires the real part of the eigenvalues of the
Fokker-Planck Hamiltonian on the complexified manifold to be
non-negative. This property, which is easily established in the real case, has
only been observed numerically for certain actions in the complex version
\cite{Klauder:1985ks}.  Occasionally, convergence to a correct representation
has been guessed from the flow of walkers on the complexified manifold and the
nature of the fixed points \cite{Haymaker:1987ey}, but this is not
conclusive.

Most studies on the validity of the complex Langevin method are based on a
mixture of analytical arguments and Monte Carlo simulations analyzing the
convergence of the algorithm to correct solutions or not (see e.g.
\cite{Aarts:2009uq,Aarts:2011ax,Nagata:2016vkn,Nishimura:2015pba}).
The present work complements those studies. Instead of the Langevin process,
our focus is on the Fokker-Planck equation, a second order differential
equation on the complexified manifold describing the evolution of $\rho(z,t)$,
and on its stationary solution, $\rho(z)$.  While a complex probability admits
many different representations, a density $\rho(z)$ projects to a unique
complex probability $P(x)$.  A natural way to justify the validity of the
complex Langevin method is to start with the Fokker-Planck equation fulfilled
by $\rho(z)$, and to try to show that its projection $P(x)$ fulfills the
projected Fokker-Planck equation (a differential equation on the real
manifold) with the target complex probability $e^{-S(x)}$ as unique
solution. This is the path followed in \cite{Salcedo:1993tj}. However, it was
shown there that the projected Fokker-Planck equation may admit further
solutions besides $e^{-S(x)}$. Moreover, the actual stochastic process may
choose one of the wrong solutions. Nevertheless, the cases analyzed in that
reference were somewhat artificial, and such spurious solutions are not often
found in practice. Here we question the premise, namely, whether the naive
projected Fokker-Planck equation is actually obtained or not after projection
from the complex to the real manifold. We find that in general, besides the
standard naive terms, additional surface terms may appear in the projected
equation, and such anomalous terms introduce a bias in the expectation value
of the observables. The crucial role played by boundary terms from integration
by parts has been noted before \cite{Aarts:2009uq}. Here we isolate the
anomaly and show in concrete examples that it is non vanishing.  Further, we
check, by numerical solution of the stationary Fokker-Planck equation, that
the various anomalous relations derived are fulfilled at a numerical level.
The origin of the anomaly is the slow falloff at infinity of $\rho(z)$ in the
imaginary direction. In order to analyze the relevant region $y=\infty$, we
introduce changes of variables which effectively compactify the complexified
manifold. The numerical solutions are obtained by solving the differential
equation in the original and in the compactified coordinates and matching both
solutions. The analysis shows that the presence of an anomaly, and hence a
bias, can be a quite general phenomenon in the complex Langevin approach.

The paper is organized as follows: Spurious solutions are discussed in
Sec. \ref{sec:spu}. The form of the anomaly is isolated in
Sec. \ref{sec:anom}. In Sec. \ref{sec:anu1} anomalous relations are derived
for a periodic action in one dimension. In Sec. \ref{sec:proof} we prove
that the anomaly is not vanishing for that  action. Generalizations
to other periodic actions are discussed in Sec. \ref{sec:3.E}. A non-periodic
action is analyzed in Sec. \ref{sec:nonp}. Finally, conclusions are presented
in Sec. \ref{sec:con}.

\section{Spurious solutions of the complex Langevin equation}
\label{sec:spu}

The complex Langevin approach is by far the most frequently used method to
construct representations of complex probabilities. The method is easy to
implement efficiently and most importantly, if the action $S(z)$ in $P=e^{-S}$
is local, i.e., only neighboring sites are coupled, so is the complex Langevin
algorithm. The main drawback is that, unlike real Langevin or other Monte
Carlo approaches for positive probabilities, the complex algorithm does not
have a sound mathematical basis. In fact, for otherwise regular complex
probabilities on the real axis, the complex Langevin process may drift to
infinity, or stabilize at a wrong solution. The fact that this cannot be
prevented is a serious shortcoming of the method as a reliable tool. Another
limitation of principle is that the method requires the target probability
$P(x)$ to have an analytical extension on the complex plane. In practice this
is not a crucial problem since in many cases of interest in physics, $S(z)$ is
holomorphic and so is $P(z)$.

As we will discuss in a moment, the complex Langevin stochastic process
dictates the evolution of a probability density $\rho(z,t)$ on the complex
plane $\C^n$ representing some complex probability $P(x,t)$ on the real
manifold $\R^n$. (Here we are simplifying; the construction can be carried out
on more general manifolds, such as Lie groups or coset spaces.)  The first
problem is whether the stochastic process converges at all. Assuming that this
is the case, the main goal is to make sure that complex probability at
equilibrium, $P(x)$, fulfills the Fokker-Planck like equation
\begin{equation}
0 = \vnabla^2 P(x) + \vnabla (\vnabla S(x) P(x))
.
\label{eq:3.1}
\end{equation}
Whether this is the case or not will be the subject of the subsequent
discussion. Momentarily we assume that \Eq{3.1} holds. One can see that $P(x)
= e^{-S(x)}$ is a solution of the equation and often this is actually the
unique solution in the space of normalizable functions. When the solution is
unique it automatically follows that the complex Langevin algorithm correctly
produces a representation of the target complex probability,
$e^{-S(x)}$. However, a caveat sometimes overlooked \cite{Lee:1994swa} is that
one has to seek solutions in the space of distributions, and not only in the
space of regular functions: because a distributional $P(x)$ can be represented
on the complex plane, the equilibrium solution $\rho(z)$ of the Langevin
process may also correspond to a distributional complex probability on the
real manifold. Without going into details, in \cite{Salcedo:1993tj} (see also
\cite{Pehlevan:2007eq}) it was shown that spurious solutions can be
constructed through the following device (here we consider a one-dimensional
problem for simplicity). For any observable $A(x)$ and path $\Gamma$ on the
complex plane, let
\begin{equation}
\esp{A}_\Gamma  = \frac{\int_\Gamma dz \, P(z) A(z)}{\int_\Gamma dz \, P(z) }
,
\label{eq:2.2b}
\end{equation}
where $P(z)= e^{-S(z)}$ is the analytical extension of the target complex
probability. This procedure defines a complex probability distribution
$P_\Gamma(x)$ through
\begin{equation}
\esp{A}_\Gamma  = \int dx \, P_\Gamma(x) A(x)
.
\end{equation}
Whenever $\Gamma$ connects two zeros (finite or infinite) of $P(z)$, or
encircles a singularity of $P(z)$ (so that $\int_\Gamma dz \, P(z) $ is not
zero), $P_\Gamma(x)$ turns out to be a solution of \Eq{3.1}. Homologous
paths define the same distribution. Concrete examples are shown in
\cite{Salcedo:1993tj}.

In practice spurious solutions are only generated by the complex Langevin
algorithm if a kernel\footnote{A kernel \cite{Okamoto:1988ru} is a
  modification in the algorithm that produces $0=-\partial_i (
  G^{ij}(x)(\partial_j P(x) + \partial_j S(x) P(x)))$. When $G_{ij}(x)$ is a
  flat metric the modified process is equivalent to one without kernel but in
  different variables.} is chosen to select them or if $P(x)$ has zeros or
singularities close to the real axis. In this regard, it should be noted that
even if the action is regular in the natural coordinates, terms of the measure
can effectively go into the action and introduce singularities there
\cite{Karsch:1985cb}.

The reason for the preference of the algorithm for smooth complex
probabilities is related to the diffusion term which tends to erase any wild
$x$ dependence. For instance, for the action $S(z) = a z^4$ with $\Re(a)>0$,
integration along the real or along the imaginary axis defines two different
complex probabilities, both fulfilling \Eq{3.1}. However, the $\rho(z)$
produced by the complex Langevin algorithm will be smooth in $x$ and the
Fourier modes $\esp{e^{-ikz}}$ will go to zero for large $k$. This corresponds
to $P(x)=e^{-a x^4}$. For the complex probability distribution defined through
integration along the imaginary axis $\esp{e^{kz}}$ will go to zero for large
$k$ but $\esp{e^{-ikz}}$ will not.

\section{Anomaly in the projected Fokker-Planck equation}
\label{sec:anom}

Wrong solutions can appear even for healthy looking complex
actions if \Eq{3.1} is not fulfilled due to the presence of anomalous
terms. This can be illustrated with one exactly solvable case, namely, that
with action\footnote{Besides a quadratic action and the trivial case
  $\rho(z)=e^{-s(x)}\delta(y)$ when $S(x)$ is real, we are not aware of further exact
  equilibrium solutions of any complex Langevin dynamics.}
\begin{equation}
S(x) = i\beta_I \cos(x)
,\qquad
\beta_I \in \R
.
\label{eq:3.5}
\end{equation}
\begin{figure}[ht]%langevin/mth/23.nb
\begin{center}
\epsfig{figure=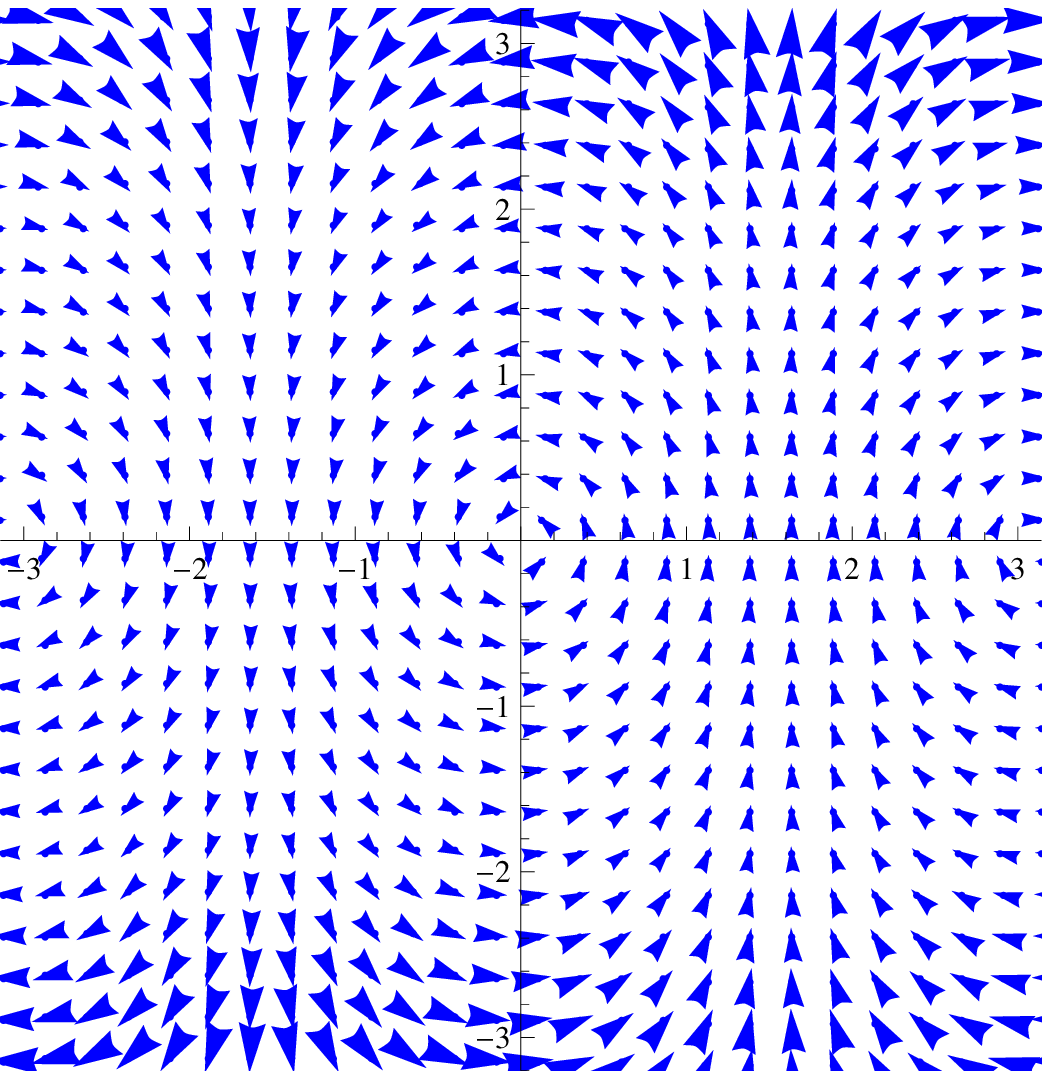,height=60mm,width=60mm}
\epsfig{figure=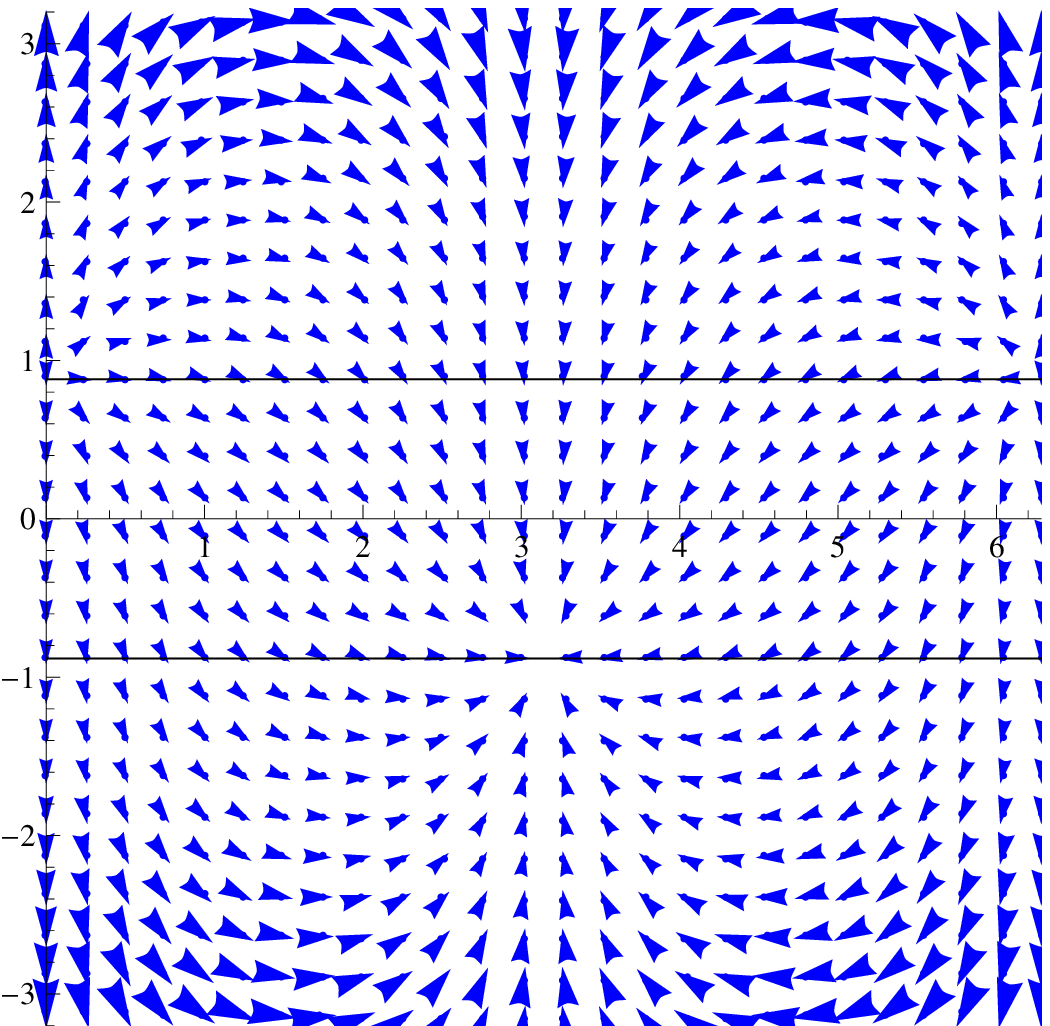,height=60mm,width=60mm}
\end{center}
\caption{Velocity fields. Left: $S=i\cos(z)$ with
  $z\in[-\pi,\pi]\times[-\pi,\pi] $.  Right: $S=\cos(z)+iz$ with
  $z\in[0,2\pi]\times[-\pi,\pi] $. The two horizontal lines at
  $y=\pm\mathrm{arcsinh}(1)$ indicate the strip where $v_y(x) \le 0$ for all
  $x$.}
\label{fig:5}
\end{figure}
Here $x$ is a periodic variable and $e^{-S(x)}$ is normalizable in
$[0,2\pi]$. The corresponding velocity flow is shown in
Fig. \ref{fig:5}$a$. For this action the Fokker-Planck equation on the complex
plane [\Eq{2.2} below] can be solved in closed form and gives\footnote{In the
  periodic case we use the normalizations $\int_0^{2\pi}dx\, P(x) =
  \int_0^{2\pi} dx \int_{-\infty}^\infty dy \, \rho(x,y)=1$.}
\begin{equation}
\rho(x,y) = \frac{1}{4\pi} \frac{1}{\cosh^2(y)}
.
\label{eq:3.6}
\end{equation}
This density is translationally invariant (with respect to $x$) so it
corresponds to $P(x)=1/(2\pi)$ and certainly is not a representation of $P(x) =
e^{-i\beta_I \cos(x)}$ unless $\beta_I=0$. Note that the expectation values
$\esp{e^{\pm iz}}_\rho$ are not ambiguous since they are absolutely
convergent, however they vanish, unlike the correct result $\esp{e^{\pm ix}} =
i J_1(\beta_I)/J_0(\beta_I)$. The correct results are reproduced by using a
representation of the two-branch type [cf. \Eq{2.102}]. The distribution
$P(x)=1/(2\pi)$ is not a solution of \Eq{3.1}, so this is not one of the
spurious solutions discussed in the previous Section.

To analyze the problem related to violations of \Eq{3.1} we will consider just
the one-dimensional case with no kernels. Since in the literature one can find
``proofs'' that the complex Langevin method must converge to the correct
solution and much of what is known relies on numerical experiments, we would
like to reach here mathematically solid conclusions at least for a simple but
non trivial case. In the concrete example considered below [\Eq{3.15}] we find
that $\esp{e^{-ikx}}$ is correctly reproduced by the complex Langevin method
when $k=-1,0,1,2,3,\ldots$ but wrong values are obtained for $k=-2,-3,\ldots$
For another study of the reliability of the complex Langevin algorithm see
\cite{Aarts:2011ax}.

Recall that complex Langevin is a Markovian process with walker $z(t)$ moving
on the complex plane according to the equation
\begin{equation}
dz(t) = v(z(t)) dt + \eta(t) \sqrt{2dt}
,
\qquad
v(z) \equiv -S^\prime(z)
.
\end{equation}
We always assume that the action can be holomorphically extended to the
complex plane, so $S(z)$ is an entire function, $S^\prime(z)$ is its
derivative, $dt$ the infinitesimal fictitious time step, and $\eta(t)$ is a
real random variable independent for each $t$ with even distribution under
$\eta \to -\eta$ and normalized such that $\esp{\eta^2} = 1$.

The real and positive probability density $\rho(z,t)$ of finding the walker at
$z$ at time $t$, obeys the following Fokker-Planck equation on the complex
plane
\begin{equation}
-\partial_t \rho = h\rho
\equiv -\partial_x^2 \rho + \partial(v \rho) + \partial^*(v^* \rho) 
=
-\partial_x^2 \rho + \vnabla \cdot (\vv \rho)
.
\label{eq:2.2}
\end{equation}
Here
\begin{equation}
\partial = \frac{1}{2}(\partial_x - i \partial_y), 
\quad
\partial^* = \frac{1}{2}(\partial_x + i \partial_y)
,
\qquad
\vv = (v_x,v_y), \quad
v = v_x+iv_y
.
\end{equation}

In the complex Langevin process the walker $z(t)$ follows a drift $v(z)$ with
a random noise $\eta$ parallel to the real axis. We will assume that the
process reaches an equilibrium distribution $\rho(z)$, that is,
\begin{equation}
\lim_{t\to+\infty} \rho(z,t) =  \rho(z)
,\qquad
h\rho(z) = 0
,
\qquad \rho \ge 0
,\qquad \int d^2z\, \rho = 1 
,
\end{equation}
for arbitrary initial $\rho(z,t_0)$. 

For the complex Langevin algorithm to work, \Eq{3.1} must be a consequence of
$h\rho=0$. To discuss this matter, let us introduce the projector operator
that relates a density $\rho(z)$ with its associated complex probability
$P(x)$ on the real axis. This projector will be denoted $K$,
\begin{equation}
P(x) = (K\rho)(x)
.
\label{eq:3.8a}
\end{equation}
The form of the operator $K$ can be obtained from the relation $\esp{A(x)}_P =
\esp{A(z)}_\rho$, for any analytic observable $A$,
\begin{equation}
\int dxdy\, \rho(x,y) A(x+iy)
= \int dx \left(\int dy\,\rho(x-iy,y)\right) A(x)
,
\end{equation}
therefore
\begin{equation}
(K\rho)(x) = \int dy\, \rho(x-iy,y)
,\qquad
\rho(x-iy,y)
\equiv e^{-iy\partial_x}\rho(x,y)
.
\label{eq:3.8}
\end{equation}

The projection involves the analytical extension of $\rho(x,y)$ as a function
of $x$ (we come back to this point below). The relation is more transparent in terms of the Fourier
modes. Although many of the considerations extend to the non compact case, to
be concrete we will consider in what follows the case of periodic $P(x)$, with
period $2\pi$:
\begin{equation}
P(x) = \frac{1}{2\pi}\sum_{k\in\Z} P_k e^{ikx}
,\qquad
\rho(x,y) = \frac{1}{2\pi}\sum_{k\in\Z} \rho_k(y) e^{ikx}
.
\end{equation}
For the projection
\begin{equation}
P_k = (K\rho)_k = \int_{-\infty}^{+\infty} dy \, e^{ky} \rho_k(y)
.
\label{eq:3.10a}
\end{equation}

Let us note that $P_k=\esp{e^{-ikz}}_\rho$, as defined from integration on the
complex plane, is often conditionally convergent. A suitable prescription is
to integrate $x$ first, and this leads to the expression in \Eq{3.10a}. Of
course, a necessary condition for the complex Langevin approach to work at all
is to produce finite expectation values. Therefore we will assume that for the
stationary solution the integrals in \Eq{3.10a} are convergent for all $k$,
and in particular
\begin{equation}
\lim_{y\to \pm\infty}  e^{ky} \rho_k(y) = 0 
\label{eq:3.10}
.
\end{equation}
If this is combined with the property 
\begin{equation}
\rho^*_k(y)= \rho_{-k}(y)
,
\end{equation}
which follows from $\rho(z)$ being real, the stronger statement obtains:
\begin{equation}
\lim_{y\to \pm\infty}  e^{|ky|} \rho_k(y) = 0 
\label{eq:3.10b}
.
\end{equation}

Next, it can be easily verified that the projection operator $K$ fulfills the
following algebraic relations
\begin{equation}
K\partial_x = \partial_x K
,\qquad
K\partial_y = i\partial_x K - 2i K \partial^*
,\qquad
K\partial = \partial_x K - K \partial^*
,\qquad
Kf(z) = f(x) K \quad\text{(for analytic $f(z)$)}
,
\end{equation}
as well as
\begin{equation}
(K\partial^* p)(x) = \frac{i}{2} p(x-iy,y)\Big|_{y=-\infty}^{y=+\infty}
.
\label{eq:3.14}
\end{equation}
[As usual $f(x)|_a^b$ stands for $f(b)-f(a)$.]

These relations can be used to project the Fokker-Planck equation,
\Eq{2.2}. This gives, on the real axis,
\begin{equation}
-\partial_t P(x,t) = H P  - \A
,
\label{eq:3.13a}
\end{equation}
with
\begin{equation}
H P  \equiv -\partial_x^2 P + \partial_x (v P) 
,\qquad
\A \equiv 2i K\partial^*(v_y\rho)
.
\label{eq:3.13}
\end{equation}

Assuming that the complex Langevin process converges for large $t$, and
provided that the {\em anomalous} term $\A$ vanishes in that limit,
the naive projected Fokker-Planck equation is recovered 
\begin{equation}
H P = 0
.
\end{equation}
This is just \Eq{3.1}.

At a {\em formal} level the anomaly $\A$ would be expected to vanish quite
generally.\footnote{In what follows, by anomaly we mean the anomaly
  corresponding to the stationary solution of the Fokker-Planck equation.}
Indeed, using \Eq{3.14}
\begin{equation}
\A(x) =
-(v_y\rho)(x-iy,y)
\Big|_{y=-\infty}^{y=+\infty}
:= \A(x)_+ - \A(x)_-.
\label{eq:2.17}
\end{equation}
This is a surface term that would vanish for a sufficiently convergent
$v_y\rho$.

The anomaly can be obtained in closed form for the action $S = i \beta_I
\cos(x)$, using the known expression of $\rho(z)$ and \Eq{2.17}. This gives
\begin{equation}
\A(x)_\pm = \pm \frac{i\beta_I}{4\pi} e^{\pm ix}
,\qquad
\A(x) =  \A(x)_+ - \A(x)_-  =\frac{i\beta_I}{2\pi} \cos(x)
.
\label{eq:3.21}
\end{equation}
The same result follows from using $P=1/(2\pi)$ and the anomalous projected
Fokker-Planck equation, \Eq{3.13a}. The relation
\begin{equation}
\A(x)_+ = -\A(-x)_-
\end{equation}
follows from parity symmetry of the action, and hence of $\rho(z)$.

%_______________________
A comment is in order related to the analytical extension implied in
$\rho(x-iy,y)$. Quite independently of the complex Langevin problem, any
sufficiently convergent function $\rho(x,y)$ on the complex plane will
produce finite expectation values for a relevant set of holomorphic
observables (sufficiently well-behaved at infinity) which includes
$e^{-ikx}$. So we expect that $\rho(x,y)$ admits a decomposition in Fourier
modes with respect to $x$, with finite components $\rho_k(y)$. The density
$\rho(x,y)$ itself can have a wild non-analytic dependence (or even have
distributional character) with respect to $x$, but still we can formally work
with $\rho(x-iy,y)$ through its Fourier modes $e^{ky}\rho_k(y)$. If the sum
over $k$ of these components does not converge (and so $\rho(x-iy,y)$ does not
exist as a function) they still define a distribution and this is enough since
in practice the projected complex probability $P(x)$ [Eqs. (\ref{eq:3.8a}) and
  (\ref{eq:3.8})] or the anomaly $\A(x)$ [\Eq{2.17}] will be needed within
integrals over $x$ weighted with some observable.  In the particular case of
$\rho(x,y)$ produced by the complex Langevin algorithm we conjecture that the
dependence on $x$ will be regular due to the smoothing effect of the
diffusion, so $\rho(x,y)$ could admit an analytical extension in this case,
even if this is not required for our analysis. What will be relevant for the
vanishing or not of the anomaly is whether $e^{ky}\rho_k(y)$ goes to zero
sufficiently fast for large $y$.
%_______________________

Before proceeding we want to make the following observation: the Fokker-Planck
equation can also be written as
\begin{equation}
-\partial_t \rho
=
-\partial^2\rho + \partial(v\rho)
-\partial^*{}^2\rho + \partial^*(v^*\rho)
-2\partial\partial^*\rho
.
\end{equation}
Upon projection to the real plane, the last term is formally zero in the sense
that for sufficiently convergent functions $K\partial^*\equiv 0$. If this
term is removed one obtains the separable solution $\rho = |P|^2$ which is
real and positive but never normalizable. $|P(z)|^2$ is formally a
representation of $P(x)$ since $\int d^2z \, A(z) P(z) P(z)^* =\int dz \, A(z)
P(z) \int dz^* \, P(z)^* = \int dz \, A(z) P(z) = \esp{A}_P$. As noted in
\cite{Salcedo:1996sa}, the convolution of a representation with a radially
symmetric positive function automatically produces a new representation, so in
principle $|P|^2$ could be transformed into a valid representation by applying
a convolution.\footnote{For instance, a Gaussian $e^{-x^2/a}$ convoluted with
  $e^{-x^2/b}$ gives $e^{-x^2/(a+b)}$; a negative $a$ can be compensated with
  a sufficiently positive $b$.} This is actually the role of the last term
$-2\partial\partial^*\rho = -\frac{1}{2}\nabla^2\rho$, which implements a
radially symmetric diffusion. The quadratic form $-\partial^2-\partial^*{}^2$
is not positive definite as it has a component of diffusion in the $x$
direction and another of anti-diffusion in the $y$ direction. The term
$-\frac{1}{2}\nabla^2\rho$ just cancels the anti-diffusion.  (One can add a
further term $-\lambda\nabla^2$, $\lambda\ge 0$ and still have an algorithm
formally equivalent to complex Langevin. The solution will be more dispersed
which in general is not what one wants.) So the complex Langevin algorithm is
essentially equivalent to sampling $|P|^2$ but introducing some (minimal)
diffusion at each step to obtain a normalizable representation. A problem with
this approach is that neither $|P|^2$ nor the radially symmetric diffusion
have a preference for the real axis as integration path, so spurious solutions
are not always screened out.

\section{Analysis of a $\U(1)$ action}
\label{sec:anu1}

In this Section we will analyze the anomaly problem for the following periodic
action
\begin{equation}
S(x) = \beta \cos(x) + i m x
,\qquad
\beta, m > 0, \quad m\in \Z
.
\label{eq:3.15}
\end{equation}
This study is of interest because it would seem that the complex Langevin
method should work for this action (see e.g. \cite{Haymaker:1987ey}). The
reason to expect this is that the deterministic flow (i.e., the velocity field
$\vv$) has an {\em attractive fixed point} at
\begin{equation}
(x=\pi,y=-y_0)
,\qquad
y_0 \equiv \mathrm{arcsinh}(m/\beta)
.
\end{equation}
(See Fig. \ref{fig:5}$b$.) This point is outside the real axis and this is a
desirable property to obtain complex expectation values for real observables
such as $\cos(x)$. Moreover, if one computes the expectation values
$\esp{e^{\pm i x}}$ for various $\beta$ and $m$ using complex Langevin they
turn out to be numerically correct.

However there are reasons for concern. For one thing, actions of the type
$\beta\cos(z)$ with complex $\beta$ behave as $\frac{1}{2}\beta e^{\pm i z}$
for $y\to \mp \infty$, so the phase of $\beta$ can be absorbed by a shift in
$x$ and only the modulus matters. This implies that real or imaginary $\beta$
are qualitatively similar in the region relevant to the anomaly, and we have
already seen that for the imaginary case the algorithm gives incorrect
results, at least for $m=0$.

For another, the flow and the complex Langevin process on the complex plane is
qualitative similar whether the real parameter $m$ is an integer number or
not. Regardless of this, there will be some $\rho(z)$ of equilibrium which
certainly will be periodic by construction. Therefore such $\rho(z)$ cannot
represent $e^{-S(x)}$, which is not a periodic function for non integer $m$.

To proceed with the analysis of the action in \Eq{3.15} we will exploit the
following observation. For an action of the type
\begin{equation}
S(x) = S_0(x) + i m x
, \qquad S_0(x) \in \R , \quad m > 0
,
\end{equation}
it follows immediately that $v_y(x,0) = -m < 0$ for all $x$. Consequently a
walker below the real axis will never cross it again. If an equilibrium
solution exists and is unique, this implies that its support will be contained
entirely on the lower half plane, $y \le 0$.

Specifically, for $S = \beta \cos(x) + i m x$ one finds that $v_y < 0$ on the
strip $|y| < y_0$ (Fig. \ref{fig:5}$b$), thus the support of the equilibrium
solution will lie in the half-plane $y \le -y_0$.
\begin{equation}
\rho=0 \quad \text{for}\quad y>-y_0
.
\end{equation}

Another useful property of this action, which holds whenever $S_0(x)$ is an
even function, is that the flow, and hence $\rho(z)$, is reflection symmetric
with respect to the imaginary axis, even if $P(x)=e^{-S(x)}$ does not have
such a symmetry,\footnote{This emergent symmetry is not the signal of a
  problem, it also appears in the two-branch representations discussed in the
  Introduction.}
\begin{equation}
\rho(x,y) = \rho(-x,y)
.
\end{equation}

A direct application of \Eq{2.17} gives for the
anomaly (noting that the upper limit $y\to +\infty$ vanishes in our case)
\begin{equation}
\A(x) =  \left( 
\frac{1}{2}\beta\cos(x)\sinh(2y)+i\beta\sin(x)\sinh^2(y) - m 
\right)
\rho(x-iy,y)\Big|_{y=-\infty}
.
\label{eq:3.18}
\end{equation}

At this point an argument can be given suggesting that the anomaly does not
vanish: In the periodic case $x$ is a compact variable, so it can be expected
that the random noise tends to flatten the $x$ dependence of $\rho$. Thus in
the large $y$ limit, the variation with $y$ will be much more important.
Neglecting $\partial_x\rho$ in the Fokker-Planck equation one obtains
(using the Cauchy-Riemann equation $\partial_x v_x = \partial_y
v_y$)\begin{equation}
0 \approx (\partial_x v_x)\rho + \partial_y(v_y \rho)
= \frac{1}{v_y}\partial_y(v_y^2 \rho)
,
\end{equation}
and hence the following estimate is obtained
\begin{equation}
\rho(x,y) \approx \frac{C(x)}{v_y^2(x,y)}
.
\label{eq:2.20}
\end{equation}
($C(x)$ is an integration constant, with respect to $y$.) Since one expects
$\rho(x-iy,y)$ to behave like $\rho(x,y)$ for large $y$ (if the Fourier mode
$k=0$ is dominant) this implies in our case
\begin{equation}
\rho(x-iy,y) \sim e^{2y} \qquad \text{for} \quad y \to -\infty 
.
\end{equation}
This asymptotic behavior allows a non vanishing anomaly in \Eq{3.18}.

Incidentally, the ansatz in \Eq{2.20} will actually be exact when $v_y$ is a
separable function of $x$ and $y$. In this case $C(x)$ can be chosen in such a
way that $\rho$ depends only on $y$. However, the additional condition requiring $v(z)$
to be an holomorphic function leaves as essentially unique solution that
already given in Eqs. (\ref{eq:3.5}) and (\ref{eq:3.6}).

The annoying analytic extension implied in $\rho(x-iy,y)$ in \Eq{3.18} can be
dealt with by using Fourier modes. Because $\rho$ is an even real function of
$x$ it follows that
\begin{equation}
\rho_k(y) = \rho_{-k}(y) = \rho_k^*(y) 
\qquad \forall y,k
.
\end{equation}
An easy calculation gives for the anomaly (\ref{eq:3.18}) in terms of Fourier
modes
\begin{equation}
\A_k = 
\left(
\frac{1}{2}\beta \sinh(y)(\rho_{k+1}+\rho_{k-1})-m\rho_k
\right) e^{ky} \Big|_{y=-\infty}
,
\end{equation}
which can be simplified using \Eq{3.10}:
\begin{equation}
\A_k = 
-\frac{1}{4} \beta  e^{(k-1)y} \rho_{k+1} \Big|_{y=-\infty}
.
\end{equation}

This result can be sharpened using \Eq{3.10b}, which results in
\begin{equation}
\A_k = 
\left\{\begin{matrix} 0 & k\ge 0 \\ 
-\frac{1}{4}\beta e^{(k-1)y} \rho_{k+1}(y)  \Big|_{y=-\infty} & k<0
\end{matrix}\right.
.
\label{eq:3.35a}
\end{equation}

The condition that $\A_k$ should take a {\em finite} value
(necessary if the complex Langevin algorithm should work at all) implies that
actually $\rho_k$ must vanish at least as $e^{(2+|k|)y}$ for large negative
$y$, i.e.,
\begin{equation}
\lim_{y\to -\infty} e^{-(2+|k|)y}\rho_k(y) < \infty
.
\label{eq:3.35}
\end{equation}
This condition is more restrictive than \Eq{3.10b}.

According to \Eq{3.35a}, the first possible anomaly comes from $k=-1$,
\begin{equation}
\A_{-1} = -\frac{1}{4}\beta e^{-2y} \rho_0(y)  \Big|_{y=-\infty} 
.
\end{equation}

In order to see the effect of the anomaly on the expectation values
$\esp{e^{-ikx}}=P_k$, let us rewrite the projected Fokker-Planck equation,
\Eq{3.13a}, in terms of Fourier modes
\begin{equation}
-\partial_t P_k
=
k(k+m) P_k -\frac{1}{2}\beta k (P_{k+1} - P_{k-1}) - \A_k
.
\label{eq:3.40}
\end{equation}
When $\partial_t P_k=0$ (equilibrium) and $\A_k=0$ (no anomaly), this
recurrence relation (removing a global factor $k$) is that of the Bessel
functions. Its unique downward solution, constrained by the conditions $P_k
\to 0$ for $k\to +\infty$ and $P_0 = 1$, is
\begin{equation}
P_k = \frac{I_{m+k} (-\beta)}{I_m(-\beta)}
.
\label{eq:3.37}
\end{equation}
These are the correct expectation values of $e^{-ikx}$ for $S(x)=\beta
\cos(x)+imx$.

We assume that, as a result of the diffusion term in the complex Langevin,
$P_k\to 0$ for large $k$. Noting that the quantities $\A_k$ vanish for $k\ge
0$ and using $P_0 = 1$, it follows from \Eq{3.40} that the $P_k$ obtained by
complex Langevin will be correct for all $k\ge -1$,
\begin{equation}
\Delta P_k = 0 \quad\text{for}\quad k\ge -1 
.
\label{eq:3.43a}
\end{equation}
($\Delta P_k$ is the shift in the complex Langevin estimate compared to the
unbiased result.)
The first anomalous (i.e., biased) expectation value takes place for
$P_{-2}$. \Eq{3.40} at equilibrium provides the following relation for $k=-1$,
\begin{equation}
\Delta P_{-2} = -\frac{2}{\beta} \A_{-1}
=
\frac{1}{2} e^{-2y} \rho_0(y)  \Big|_{y=-\infty} 
.
\label{eq:3.43}
\end{equation}

\section{Proof of the presence of anomalies}
\label{sec:proof}

\subsection{Compactification and numerical solutions}

From inspection of the Fokker-Planck equation or the complex Langevin process
is not easy to decide whether $\rho_0(y)$ goes to zero faster than $e^{2y}$
for large negative $y$ or not, as required in \Eq{3.43}. In order to clarify
this issue we will make a change of variables, from $(x,y)$ to $(X,Y)$, or
$(R,\varphi)$ in polar form, such that $\varphi \equiv x$ and $y= -\infty$ will be
the new origin $R=0$,
\begin{equation}
\varphi =x ,\qquad 
R = -\frac{1}{\sinh(y)} \quad (y<0)
,
\qquad
X = R \cos(\varphi)
,
\quad
Y = R \sin(\varphi)
.
\label{eq:3.41}
\end{equation}
The coordinates $(X,Y)$ only cover the lower half complex plane of $z$. This
is sufficient in our case. The original manifold was a cylinder due to
$x=0\equiv 2\pi$. In the new coordinates we have compactified the lower end of
this cylinder, $y=-\infty$, to a point, $R=0$, which is now a regular point of
the manifold $(X,Y)$ (namely, its origin). We want to show that $R=0$ is also
a regular point of the complex Langevin process in the new variables.

Let us denote $\sigma$ the new density in coordinates $(X,Y)$
\begin{equation}
\int_0^{2\pi} dx \int_{-\infty}^0 dy \, \rho = \int_{\R^2} dX \, dY\, \sigma
,
\qquad
\rho = \chi R^2 \sigma
,
\qquad
\chi \equiv \sqrt{1+R^2}
.
\end{equation}
We introduce Fourier modes as usual
\begin{equation}
\sigma = \frac{1}{2\pi}\sum_k e^{ik\varphi} \sigma_k(R)
,\qquad
\sigma_k = \sigma_{-k} = \frac{1}{\chi R^2} \rho_k 
.
\end{equation}

The radial coordinate $R(y)$ is such that $\rho \asymp 4 e^{2y}\sigma$ for
large and negative $y$, thus
\begin{equation}
\frac{1}{8\pi}\rho_0(y)e^{-2y} \big|_{y=-\infty} = \frac{\sigma_0(0)}{2\pi} =
 \sigma\big|_{R=0} \equiv  \sigma(0) .
\end{equation}
In this way the leading anomaly, \Eq{3.43}, becomes
\begin{equation}
\Delta P_{-2} = 4\pi \sigma(0)
,
\label{eq:3.44}
\end{equation}
and its value depends on the value of the density $\sigma(X,Y)$ at the origin.

In the new coordinates, the Fokker-Planck equation takes the form
\begin{equation}
-\partial_t\sigma = -\partial_\varphi^2 \sigma + \partial_X (v_X \chi \sigma)
+ \partial_Y (v_Y \chi \sigma)
\label{eq:2.35}
\end{equation}
with
\begin{equation}
v_X = - m X - \beta 
,
\qquad
v_Y =  - m Y 
,\qquad
\partial_\varphi = X\partial_Y - Y \partial_X
.
\label{eq:3.48}
\end{equation}
The most relevant condition on the choice of $R(y)$ is that $R\sim e^y$. This
ensures that $(v_X,v_Y)$ is finite at $y\to -\infty$.  The precise form of
$R(y)$ in \Eq{3.41} is such that $v_Y$ has no contribution from $\beta$.  The
function $\chi$ is harmless since equals $1$ at the origin and is everywhere
smooth (in fact analytic) on the $(X,Y)$ plane. For the present discussion
$\chi$ can be absorbed in $\sigma$.

\begin{figure}[ht]
\begin{center}
\epsfig{figure=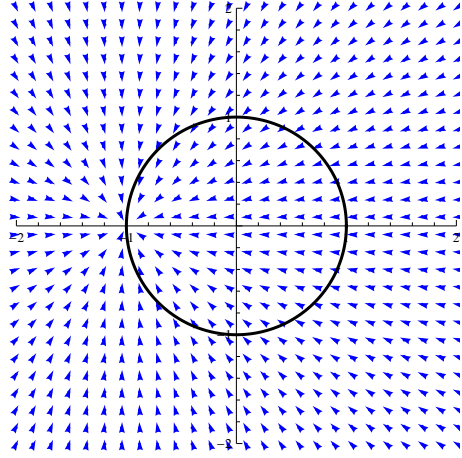,height=60mm,width=60mm}
\end{center}
\caption{Velocity field of $S=\cos(z)+iz$ in variables $(X,Y)$ in the range
  $[-2,2]\times[-2,2]$. The fixed point is at $(-1,0)$. The circle indicates
  the boundary of the support of the equilibrium solution, i.e. the locus of
  $y=-y_0$. The diffusion moves the walker along circles centered at the
  origin (not displayed in the figure).}
\label{fig:7}
\end{figure}
The flow $(v_X,v_Y)$ is displayed in Fig. \ref{fig:7}.  We can see from
\Eq{3.48} that in $(X,Y)$ coordinates the deterministic part of the flow is
just an inwards radial field with center at the attractive fixed point
$(X=-R_m,Y=0)$, with $R_m\equiv \beta/m$. The point $R=0$ (corresponding to
$y=-\infty$) is not special in any way for that flow.  On top of this,
$-\partial_\varphi^2\sigma$ is an angular diffusion term which accounts for
random shifts in the angle $\varphi$ of the walker without changing $R$. So
this diffusion term does not directly produce an increase nor decrease of the
density $\sigma$ at the origin.

The stochastic angular jumps are of order $1$ (times $\sqrt{dt}$) in the
variable $\varphi$ but they are small in the $(X,Y)$ variables as $R$
approaches zero. The fields $(v_X,v_Y)$ and $\partial_\varphi$ are regular
everywhere.  In the $(X,Y)$ variables $R=0$ is a regular point of the
stochastic process and there is no mechanism at work that would enforce
$\sigma(0)=0$.  The conclusion is that $\sigma(0)>0$ and there is a bias in
the complex Langevin method already for $P_{-2}$.

\begin{figure}[ht]%langevin/mth/36.nb
\begin{center}
\epsfig{figure=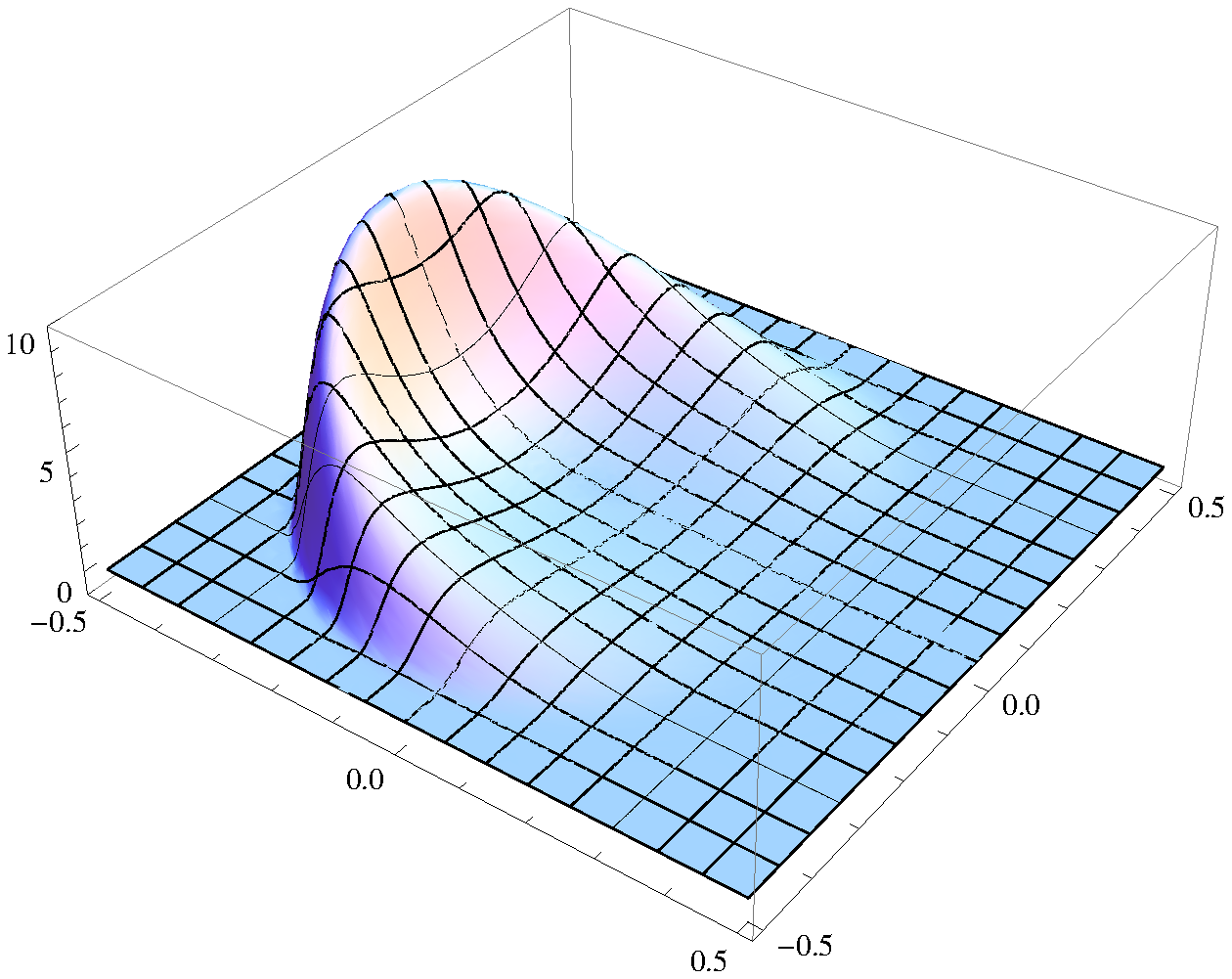,height=70mm,width=80mm}
\epsfig{figure=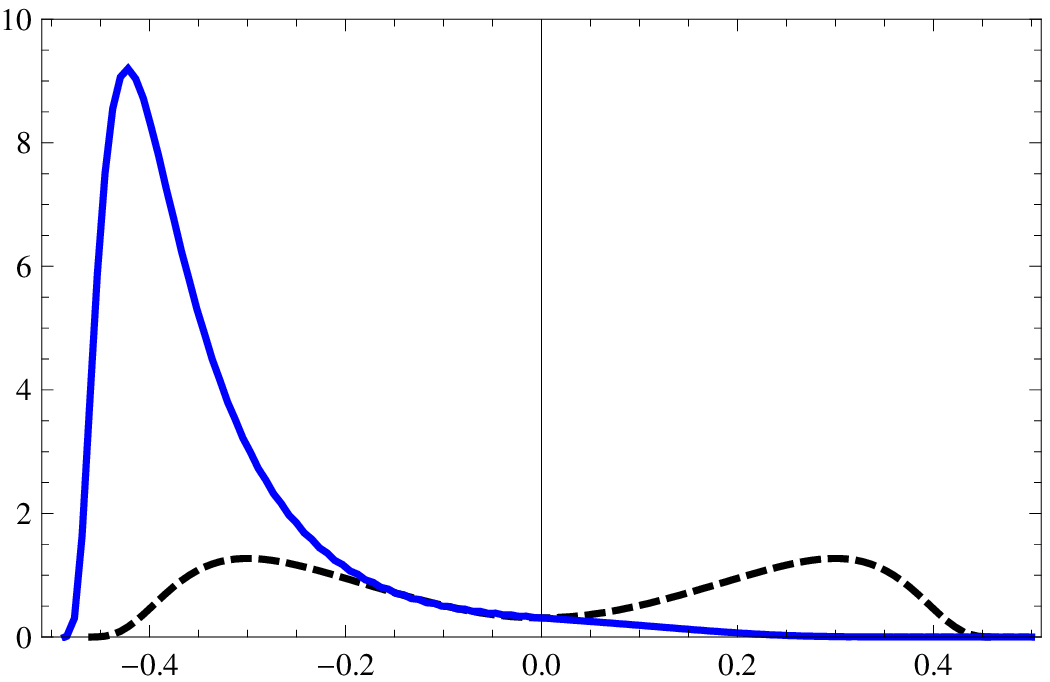,height=50mm,width=80mm}
\end{center}
\caption{Function $\sigma(X,Y)$ for $\beta=0.5$ and $m=1$. Left: $\sigma$ on
  the $(X,Y)$ plane. The support is located on $R\le R_m=\beta/m$. The
  numerical calculation uses $64$ Fourier modes in $\varphi$ and $64$ points in the mesh of
  the variable $R$. Numerical derivatives in $R$ were obtained using 5 points.
  Right: Sections of $\sigma(X,Y)$ along the $X$ (solid blue line) and $Y$
  (dashed black line) axes. $\sigma(0)=0.310$.}
\label{fig:2}
\end{figure}
The conclusion just mentioned is confirmed by numerical solutions of \Eq{2.35}
for the stationary case (see Fig. \ref{fig:2}).  Among the various numerical
approaches used (including polar and Cartesian coordinates), we have obtained
best result by using Fourier modes for the dependence on the angular variable
$\varphi$ (rather than directly a mesh on $\varphi$) and a regular grid for
$R$. For the solution displayed in Fig. \ref{fig:2} we have used $64$ Fourier
modes for $\varphi$ and $64$ points in $R$.  We have carried out calculations
for several values of $\beta$ and $m$. As it would be expected, the value of
$\sigma(0)$, and hence the anomaly, is enhanced for small values of $\beta$ or
large values of $m$.

The expectation values $P_k$ numerically obtained from this $\sigma(X,Y)$
agree well with the Bessel function exact result for $k=+1,-1,2$. For $k=-2$,
the exact expectation value is $P_{-2} = 1$. Instead of this, numerically one
obtains $P_{-2}=4.86$, for $\beta=0.5$ and $m=1$. This bias is quite
consistent with the relation (\ref{eq:3.44}),
\begin{equation}
\Delta P_{-2}=3.86
,
\qquad
4\pi\sigma(0)=3.89
\qquad (\beta=0.5, \ m=1 )
\,.
\end{equation}
A standard complex Langevin simulation also correctly reproduces $P_k$ for
$k=+1,-1,2$ but becomes noisy for $k=-2$.

As we mentioned before, the unphysical case of non integer $m$ is
qualitatively similar to that of integer $m$ as regards to the complex
Langevin process on the complex plane, and this is also obvious from
\Eq{3.48}. Remarkably, the arguments leading to Eqs. (\ref{eq:3.43a}) and
(\ref{eq:3.43}) hold for non integer $m$ as well and we have confirmed this
numerically: For $m=0.5$ or $m=\sqrt{2}$ we still find that the $P_k$ as
obtained from $\sigma(X,Y)$ reproduce \Eq{3.37} for $k\ge -1$ whereas $P_{-2}$
fulfills (\ref{eq:3.44}) at a numerical level.

\subsection{Proof of the presence of anomalies from analyticity}

Because establishing in an unambiguous manner the existence of a bias in the
expectation values of $e^{-ikx}$ is important, we provide a different
argument, based on analyticity of $\sigma(X,Y)$. In the absence of angular
diffusion, the walkers would accumulate at the attractive fixed point
$(X=-R_m,Y=0)$ and, at equilibrium, $\sigma$ would become a Dirac delta
there. The diffusion moves the walkers along circumferences centered at the
origin. The result (see Fig. \ref{fig:2}) is an equilibrium distribution with
support on the disk $R\le R_m$. The coefficients of the differential equation
\Eq{2.35} are real-analytic functions at all points, so one should expect that
$\sigma$ is real-analytic for $R < R_m$. Certainly it cannot be analytic on
$R=R_m$ because $\sigma$ becomes identically zero for $R > R_m$ but $\sigma$
should be $C^\infty$ there (except at the fixed point) since the vanishing of
the function is not imposed as a boundary condition, instead it follows
automatically from the differential equation (this is also observed
numerically). The only point where $\sigma$ may be non smooth is at the fixed
point. There, the vanishing of $v_X$ and $v_Y$ allows to have a discontinuous
radial derivative of $\sigma$ (although the function itself is
continuous). The angular derivative must be continuous due to the smoothing
effect of the diffusion. At equilibrium $\sigma$ fulfills the differential
equation at all points except at the fixed point. As a 
distribution $\sigma$ fulfills the differential equation everywhere.

Therefore, the origin is a regular point of $\sigma(X,Y)$ and this can be used
to show that not all the anomalies $\A_k$ can vanish
simultaneously. First note that the non-trivial anomalies, \Eq{3.35a}, can be
expressed as
\begin{equation}
\A_{-k-1} = 
- \beta \left(\frac{2}{R}\right)^k \sigma_k(R)\Big|_{R=0} ,\qquad
k\ge 0
.
\end{equation}
That $\sigma$, or equivalently $\chi\sigma$, is real-analytic at $R=0$ implies
that
\begin{equation}
\chi \sigma_k =  R^{|k|} f_k(R^2)
,
\end{equation}
where the functions $f_k(\xi)$ are also analytic. Thus
\begin{equation}
\A_{-k-1} = - \beta 2^k f_k(0)
,\qquad
k\ge 0
.
\end{equation}
Thus analyticity of $\sigma$ already guarantees that the anomalies $\A_k$ are
finite (instead of divergent).

Let us assume that all anomalies were vanishing. In this case $f_k(0)=0$ for
all $k$ (recall that $f_k=f_{-k}$ since $\sigma_k$ and $\rho_k$ have the same
property). This merely implies that $\sigma=R^2\tau$ with $\tau(X,Y)$
analytic, but it is not inconsistent with $\sigma$ being analytic. To find a
contradiction one must resort to the equilibrium Fokker-Planck equation. In
terms of the $f_k(\xi)$ ($\xi \equiv R^2$) the equation takes the form
\begin{equation}\begin{split}
0 &= -2 m (f_0 + \xi f^\prime_0) - 2 \beta (f_1 + \xi f_1^\prime)
,
\\
0 &= \left(\frac{k^2}{\chi} - m(k+2) \right) f_k
-\beta(k+1)f_{k+1} - 2 m \xi f_k^\prime
-\beta (f_{k-1}^\prime +\xi f_{k+1}^\prime )
\qquad (k\ge 1)
,
\label{eq:3.55}
\end{split}\end{equation}
where the prime indicates differentiation with respect to $\xi$.

Incidentally, the first equation in (\ref{eq:3.55}), corresponding to $k=0$,
is equivalent to the condition that at equilibrium no net flux traverses any
$y=\text{constant}$ line, $\int dx v_y\rho = 0$. That equation admits the
closed solution $\sigma_1 = -\frac{m}{\beta} R \sigma_0$. Since
$|\sigma_k|\le\sigma_0$ one would obtain a contradiction for $R>R_m$. The
resolution is that $\sigma \equiv 0$ for $R>R_m$.

Coming back to the proof, let us assume that $f_k(0)=0$ for all $k$ (no
anomaly). Setting $\xi=0$ in the second equation in (\ref{eq:3.55})
immediately implies $f^\prime_k(0)=0$ for all $k$ as well.  From this, taking
a derivative with respect to $\xi$ and setting $\xi=0$ one obtains in turn
$f^{\prime\prime}_k(0)=0$.  In fact, by induction, having already
$f^{(n)}_k(0)=0$ for $0 \le n \le N$, it follows that $\partial_\xi^N(\xi
f_k^\prime)(0)=0$, so applying $\partial_\xi^N$ to the equation and taking
$\xi=0$ one concludes $f^{(N+1)}_k(0)=0$. Now, if all the derivatives of the
$f_k(\xi)$ vanish at a point and the functions are analytical they must be
identically zero, along with $\sigma$ for all $(X,Y)$. This incorrect
conclusion is avoided if some of the anomalies are not zero.

\subsection{Alternative proof of the presence of anomalies}
\label{sec:appA}

We have just shown for $S=\beta\cos(x)+imx$ ($\beta\in\R$) that the
analyticity of $\sigma(X,Y)$ plus the assumption of a vanishing of all
anomalies would imply the absurd conclusion $\sigma(X,Y)\equiv 0$. Here we
want to show that a contradiction can be obtained assuming only that all the
Fourier modes $\rho_k(y)$ fall off exponentially for large negative $y$. This
is a weaker assumption because analyticity requires positive integer powers of
$e^y$ and here we allow fractional powers.

Specifically we assume 
\begin{equation}
\forall k\ge 0 
\qquad
\lim_{y\to - \infty}  e^{-\alpha_k y} \rho_k(y) =  a_k 
,
\qquad
\lim_{y\to - \infty}  e^{-\alpha_k y} \rho^\prime_k(y) =  a_k \alpha_k 
,
\end{equation}
(the prime indicates derivative with respect to $y$) with
\begin{equation}
\alpha_k = k+2 +\delta_k,
\qquad
\delta_k \ge 0
,\qquad
a_k \not=0
.
\end{equation}
The condition $\delta_k \ge 0$ implements \Eq{3.35}. The corresponding anomaly
vanishes if $\delta_k>0$.

The Fokker-Planck equation in terms of Fourier modes of $\rho$ takes the form
\begin{equation}
0 = k^2\rho_k -m\rho^\prime_k
+ \frac{\beta}{2}  \cosh(y) \big( (k+1)\rho_{k-1} - (k-1) \rho_{k+1} \big)
+ \frac{\beta}{2} \sinh(y) \left( \rho^\prime_{k-1} + \rho^\prime_{k+1} \right)
.
\end{equation}
Using $\rho_k \sim a_k e^{\alpha_k y}$ and retaining the leading terms yields,
for $k\ge 1$,
\begin{equation}
0 = (k^2 -m(k+2 + \delta_k)) a_k e^{(k+2+\delta_k)y} 
- \frac{\beta}{4} a_{k-1} \delta_{k-1} e^{(k+\delta_{k-1})y} 
- \frac{\beta}{4}(2k+2+\delta_{k+1}) a_{k+1} e^{(k+2+\delta_{k+1})y}
+ \text{subleading}
.
\end{equation}

If we assume that all $\delta_k>0$ (no anomalies), it follows that the term
with $a_{k-1}$ does not vanish (because $\delta_{k-1}\not=0$) and this term
will be dominant, thus violating the equation, unless $\delta_{k-1}$ is
sufficiently large to match the two other terms. Specifically
\begin{equation}
\delta_{k-1} \ge 2 + \min(\delta_k,\delta_{k+1}) \quad \forall k>0
.
\label{eq:A.1}
\end{equation}
However it easy to see that such strict condition is unattainable. Starting
from some large $N$, the assumption $\delta_N,\delta_{N-1}>0$ implies
$\delta_{N-2}>2$, in turn this implies $\delta_{N-3}>2$ and hence
$\delta_{N-4}>4$, and eventually $\delta_{N-2n} > 2n$ whenever $2n\le N$. But
$N$ is arbitrary, thus \Eq{A.1} can only hold if $\delta_k$ is larger than any
even number, i.e., $\delta_k=\infty$.  In other words, any exponential falloff
is incompatible with the assumption of no anomalies. Since a
falloff faster than exponential seems incompatible with
the complex Langevin algorithm we conclude that \Eq{3.1} gets an anomaly
for the action $S=\beta \cos(x)+imx$.

The behavior $\rho \sim e^{2y}$ not only gives rise to anomalies in \Eq{3.1},
it also entails that the integrals $\int d^2z \, \rho(z) e^{-ikz}$ are not
absolutely convergent for $|k|\ge 2$, although they can be given a natural
meaning by expressing them in terms of Fourier modes. The conditional
convergence implies that the variance would diverge in a straight Monte Carlo
approach of those expectation values.  Once again we point out that there is
nothing pathological with the action $\beta\cos(x)+im$ itself; any holomorphic
observable on the finite complex plane has an absolutely convergent
expectation value using a two-branch representation of the type described in
the Introduction, since the support of such representations is bounded.

\section{Generalization to other $\U(1)$ actions}
\label{sec:3.E}

\subsection{Action \ $\beta\cos(x)$ \ with complex $\beta$}

As for generalizations of the previous analysis to other actions, let
\begin{equation}
S(x) = \beta_1 e^{ix} + \beta_{-1} e^{-ix} +imx
,\qquad \beta_{\pm1}\in\C ,\quad m\in\Z
.
\label{eq:3.53}
\end{equation}

Following similar steps as for $S = \beta \cos(x) +i m x$ one easily obtains
(using \Eq{3.10})
\begin{equation}
\A_k = -\frac{1}{2}\beta_1^* e^{(k-1)y} \rho_{k+1}(y) \big|_{y=-\infty}
-\frac{1}{2}\beta_{-1}^*  e^{(k+1)y} \rho_{k-1}(y) \big|_{y=+\infty}
.
\label{eq:3.53a}
\end{equation}
$\A_0$ must vanish since this is equivalent to conserving the number of
walkers and we have already assumed that the process has a stable normalizable
equilibrium solution.  For the remaining anomalies, in the present case there
are no symmetries helping to remove some of them. In particular barring
accidental cancellations there will be anomalies $\A_{\pm1}$ coming from
$\rho_0(y)$ at $y=\pm\infty$. In fact, doing changes of variables $R\sim
e^{\mp y}$ to compactify $y=\pm\infty$ as before, one finds again that the
points at infinity are regular, the density in the transformed coordinates are
finite (non zero) and these anomalies do not vanish. These two anomalies
contaminate all the expectation values through the recurrence relation. The
action $S = i\beta_I \cos(x)$ illustrates this effect. There all the $\A_k$
vanish except $\A_{\pm 1}$ (due to $\rho_k= 0 $ for $k\not=0$) and all
expectation values other than $P_0$ are incorrect (they vanish).

In order to sustain the previous arguments more quantitatively, let us consider
an action of the type
\begin{equation}
S(x) = \beta\cos(x)
,
\qquad \beta\in\C
.
\end{equation}
This action is an even function of $x$, correspondingly the equilibrium
solution of the Fokker-Planck equation, $\rho(z)$, is an even function of $z$.

We have proceeded to numerically solve the Fokker-Planck equation at
equilibrium. The differential equation can be written in variable $z=x+iy$,
nevertheless, in order to study the region of large $|y|$, relevant for
anomalies, it is convenient to use additional suitable
variables. Specifically, we use the variables $(X,Y)$ introduced in \Eq{3.41},
and the associated density $\sigma(X,Y)$. These variables compactify the
complex plane and have the virtue that the corresponding drift has a finite
velocity at $X=Y=0$ with vanishing diffusion there.  The variables $(X,Y)$
only cover the lower half-plane $y<0$, however, since $\rho(z)$ is an even
function we can work on that lower half-plane without loss of generality.

Our approach has been to solve the differential equation for $\rho(x,y)$ on
the strip $-y_c < y < y_c$ (actually $-y_c < y < 0$ suffices due to the
symmetry) and solve for $\sigma(X,Y)$ on the disk $R<R_c$, where
$R_c=1/\sinh(y_c)$. The two solutions are matched at $R=R_c$.  The matching
parameter $y_c$ is chosen for a given $\beta$ attending to numerical
convenience. Once again we use Fourier modes for the dependence on variable
the $x \equiv \varphi$, in both sectors, and regular meshes for $y$ and $R$. The correct
boundary condition at $R=0$ is selected by imposing that no net flux passes
through the origin (or in fact any circle $R=\text{constant}$, we take
$R=R_c$).

The exact solution is known for purely imaginary non vanishing $\beta$,
namely,
\begin{equation}
\rho = \frac{1}{4\pi} \frac{1}{\cosh^2(y)}
,
\qquad
\sigma = \frac{1}{4\pi} \frac{1}{\chi^3(R)}
\qquad
(\beta=-\beta^*)
.
\label{eq:6.4}
\end{equation}
($\chi(R)\equiv\sqrt{1+R^2}$.) Note that the integral of $\sigma$ over the
$(X,Y)$ plane is $1/2$ since this covers only $y<0$.

\begin{figure}[ht]
\begin{center}
\epsfig{figure=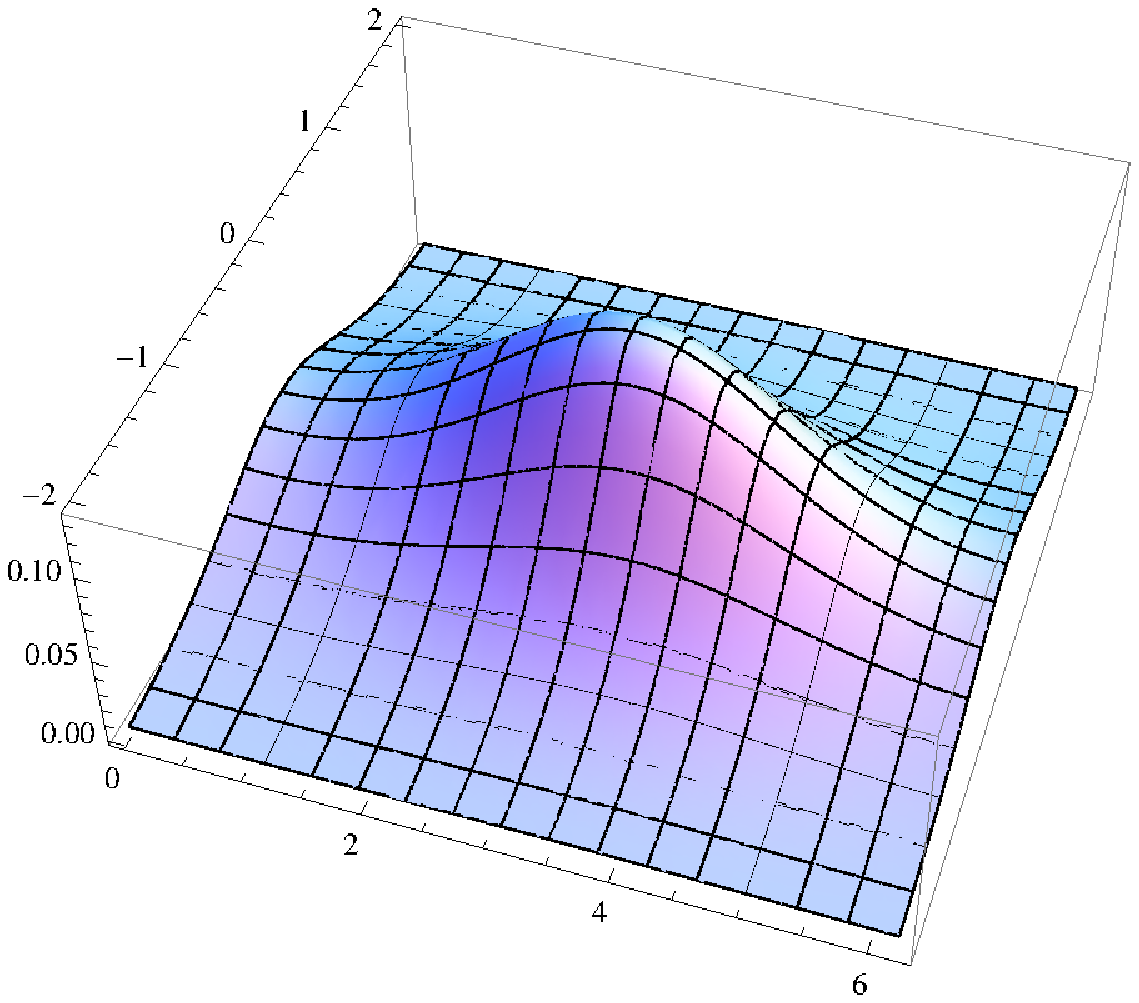,height=70mm,width=80mm}
\epsfig{figure=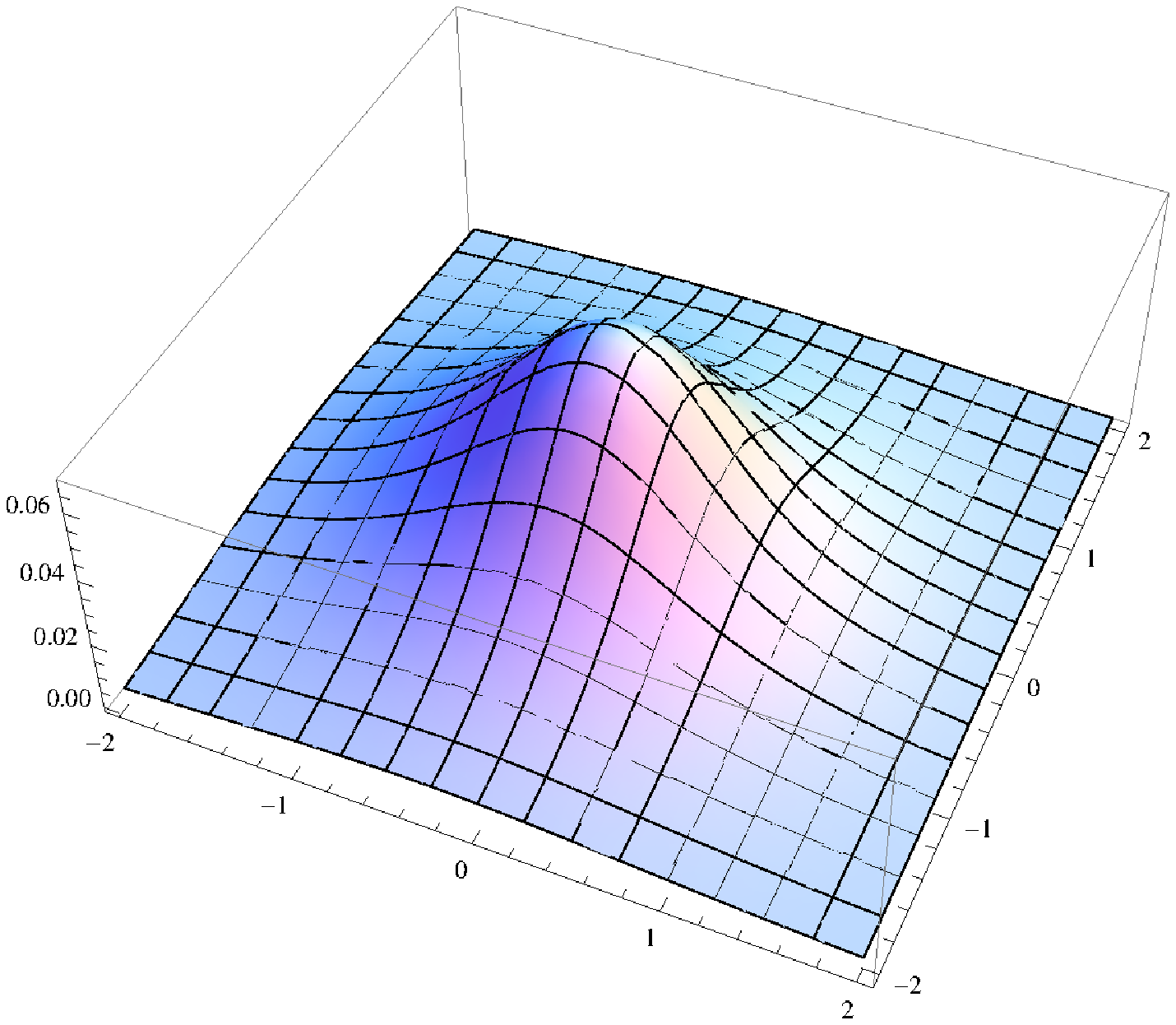,height=70mm,width=80mm}
\end{center}
\caption{Stationary solution of the Fokker-Planck equation of $S=\beta\cos(x)$
  for $\beta=0.25+i0.5$.  Left: $\rho(x,y)$ for $0\le x \le 2\pi$. This
  function is periodic in $x$ and $\rho(x,y)=\rho(-x,-y)$.  Right:
  $\sigma(X,Y)$.  In the numerical solution the matching was set at
  $R_c=2$. For $x\equiv\varphi$, $64$ Fourier modes were used and $64$ points in the $R$ and $y$
  grids.  Numerical derivatives in $R$ and $y$ were obtained using 5 points.}
\label{fig:8}
\end{figure}
A numerical solution for $\beta=0.25+i0.5$ is displayed in Fig. \ref{fig:8}.
The matching of the two sectors has been chosen at $R_c=2$. The two forms
$\rho(x,y)$ (left) and $\sigma(X,Y)$ (right) are shown. As one would
anticipate, for this $\beta$ the value of $\rho(0)$ is larger than $1/(4\pi)
\approx 0.08$ while $\sigma(0) < 1/(4\pi)$. The value $1/(4\pi)$ corresponds
to purely imaginary $\beta$ [see \Eq{6.4}]. The presence of a real part in
$\beta$ brings the distribution closer to the real axis, hence enhancing
$\rho(0)$ and quenching $\sigma(0)$.  As shown subsequently, \Eq{3.66}, a non
zero value of $\sigma(0)$, as clearly displayed in Fig. \ref{fig:8}$b$,
implies the presence of an anomaly in the projected Fokker-Planck equation.

Regarding the anomaly for the actions $\beta\cos(x)$, since the Fokker-Planck
equation preserves the parity of the action, one has
\begin{equation}
\A_k = \A_{-k}
.
\end{equation}
From \Eq{3.53a} and \Eq{3.10b}, it follows that
\begin{equation}
\A_1 = -\frac{1}{4}\beta^*e^{2y} \rho_0(y)\Big|_{y=+\infty}
=
-\beta^* \sigma_0(0)
=
-2\pi\beta^* \sigma(0)
.
\label{eq:3.66}
\end{equation}
Therefore, as expected there is an anomaly whenever $\sigma(X,Y)$ is sizable
near $R=0$. This is a regular point of the stochastic process in the new
variables and $\sigma(0)$ is non vanishing in general (Fig. \ref{fig:8}).

To see the bias introduced by the anomaly of \Eq{3.66} on the expectation
values, we again make use of the projected Fokker-Planck equation, \Eq{3.40},
with $m=0$ and complex $\beta$. Selecting the stationary case, $\partial_t
P_k=0$, gives for $k=1$, using $P_0=1$,
\begin{equation}
0
=
P_1 - \frac{\beta}{2} (P_2-1) + 2\pi\beta^* \sigma(0)
.
\end{equation}
This equation holds with $\sigma(0)=0$ for the exact expectation values in
\Eq{3.37}.  Therefore the biases $\Delta P_1$, $\Delta P_2$ introduced by the
anomaly fulfill the relation
\begin{equation}
\Delta P_1 - \frac{\beta}{2} \Delta P_2 = - 2\pi\beta^* \sigma(0)
.
\label{eq:6.8}
\end{equation}
Numerically, we obtain $(\beta=0.25+i0.5 )$
\begin{equation}\begin{split}
\Delta P_1 - \frac{\beta}{2} \Delta P_2
&= -0.103539 + i 0.207078
,
\\
- 2\pi\beta^* \sigma(0)
&= -0.103545 + i 0.207089
. 
\end{split}
\label{eq:6.9}
\end{equation}

%____________________
The existence of a bias for complex $\beta$ is consistent with similar
findings in the recent work \cite{Makino:2015ooa} which studies the analytically
solvable two-dimensional Yang-Mills theory with a complex coupling constant.
%_______________________

\subsection{Other $\U(1)$ actions}

One can consider more general actions of the form
\begin{equation}
S(x) = \sum_{k=n^\prime}^n \beta_k e^{ikx} + i mx,
\qquad n^\prime \le n
.
\end{equation}
The case $n=n^\prime=0$ is of no interest (and it is not normalizable unless
$m=0$) so either $n>0$ or $n^\prime<0$ or both. Let us assume $n>0$ for
definiteness. In this case the analytically extended action $S(z)$ will be
dominated by the mode $k=n$ for large negative $y$, i.e., as regards to its
behavior at $y=-\infty$, this action is equivalent to $\beta_n e^{inz}$. This
is a periodic action with period $2\pi/n$, and in each period it is equivalent
to the action $\beta_n e^{iz}$ with $z\in[0,2\pi]$, which brings us to
\Eq{3.53} already studied. Therefore, barring accidental cancellations one
should expect that anomalies are generated at $y=-\infty$ yielding a bias in
the expectation values.

For a different generalization, one can consider a lattice with variables at
sites or links and contributions of the type $\beta\cos(\phi)$ to the
action, with a chemical potential or other mechanism making the
variables to go to the complex plane. One can expect that the use of the
complex Langevin algorithm will introduce a bias in the expectation value of
the observables. The reason is that once the ensemble has reached the
equilibrium one can always keep updating the configurations without changing
that equilibrium. This is true if one chooses to update just a single
variable, and this leads us to the one-dimensional case we have been
studying. In this view, a bias in the conditional distribution of a single variable implies
a bias in the full distribution. On the other hand, updating just one variable
is not the prescription of the standard complex Langevin algorithm, so this
argument is not conclusive.

\section{Analysis of a non periodic action}
\label{sec:nonp}

We have already noted in the Introduction that a harmless looking action such
as $S= x^4/8+2ix$ cannot be reproduced by complex Langevin algorithm. The
reason is that $\esp{e^{-ix}}=-4.98$, yet at equilibrium all Langevin walkers
are below the real axis, so $|\esp{e^{-ix}}_{\rm CL}|\le 1$.

In this section we want to study the presence of anomalies in a non periodic
system. Specifically we consider the following one-dimensional action
\begin{equation}
S(x) = \beta \frac{x^4}{4}
,
\qquad  \Re\beta > 0.
\label{eq:3.100}
\end{equation}
Clearly, the symmetry $S(x)=S(-x)$ will be shared by the complex Langevin
equilibrium solution on the complex plane, 
\begin{equation}
\rho(z) = \rho(-z)
,\qquad z\in\C
.
\end{equation}
The fixed point of the deterministic flow is at $z=0$. It is attractive in a
direction (the real axis rotated by $\beta^{1/2}$) and repulsive in the
orthogonal one.\footnote{Stationary points ($S^\prime_0=0$) which are non
  degenerated ($S^{\prime\prime}_0\not=0$), are either attractive
  [$\Re(S^{\prime\prime}_0)>0$], repulsive [$\Re(S^{\prime\prime}_0)<0$], or
  neutral [$\Re(S^{\prime\prime}_0)=0$], however, $z=0$ is a degenerated fixed
  point in our case.}

For convenience, we will allow an extra isotropic diffusion term in the
complex Langevin process:
\begin{equation}
-\partial_t \rho =
-\partial_x^2 \rho + \vnabla \cdot (\vv \rho)
- \lambda \vnabla^2 \rho
,
\qquad
\lambda \ge 0
.
\label{eq:2.2a}
\end{equation}
As noted before, this is formally correct and the algorithm with non vanishing
$\lambda$ has as much mathematical justification as the standard
treatment. However the extra diffusion tends to spread the equilibrium
distribution, thereby increasing the variance in the Monte Carlo calculation
of the observables. Also a possible anomaly is enhanced as regions away from
the real axis become more populated. Mathematically, the equilibrium solution
is better behaved for positive $\lambda$, in particular, we expect $\rho(z)$
to be real analytic for all finite $z$.

For the action in \Eq{3.100}, the Fokker-Planck equation, with $\lambda$,
becomes
\begin{equation}
-\partial_t \rho =
-\beta \partial(z^3\rho) - \beta^* \partial^*(z^*{}^3\rho)
-\partial^2\rho - \partial^*{}^2\rho
-2(1 + 2\lambda) \partial\partial^* \rho
.
\end{equation}

Because the anomaly is related to the large $y$ region, we will introduce a new
set of coordinates $(X,Y)$ centered at $z=\infty$. Specifically,
\begin{equation}
Z=X+iY = \frac{1}{z^2}, \qquad z=x+iy .
\label{eq:7.5}
\end{equation}
Let us denote by $\sigma(X,Y)$ the density in the new coordinates,
\begin{equation}
\rho(z) = 4R^3 \sigma(Z),
\qquad
R = |Z|.
\label{eq:3.76}
\end{equation}

Note that the $Z$ plane only covers a half-plane of $z$. This is sufficient
due to the symmetry $z\to -z$ of $\rho(z)$. Of course, $\sigma$ is normalized
on the Riemann surface, so it has normalization $1/2$ on the $Z$ plane.

The new coordinate has been chosen so that the new deterministic flow has a
finite velocity near $Z=0$. Indeed, the Fokker-Planck equation in coordinates
$(X,Y)$ takes the form
\begin{equation}
-\partial_t \sigma =
 \partial((2\beta + 6 Z^2)\sigma) +  \partial^*((2\beta^* + 6 Z^*{}^2)\sigma)
-4\partial^2(Z^3\sigma) - 4\partial^*{}^2(Z^*{}^3\sigma)
-8(1 + 2\lambda) \partial\partial^* (R^3 \sigma)
,
\end{equation}
and the velocity at $Z=0$ is $2\beta$.

\begin{figure}[ht]
\begin{center}
\epsfig{figure=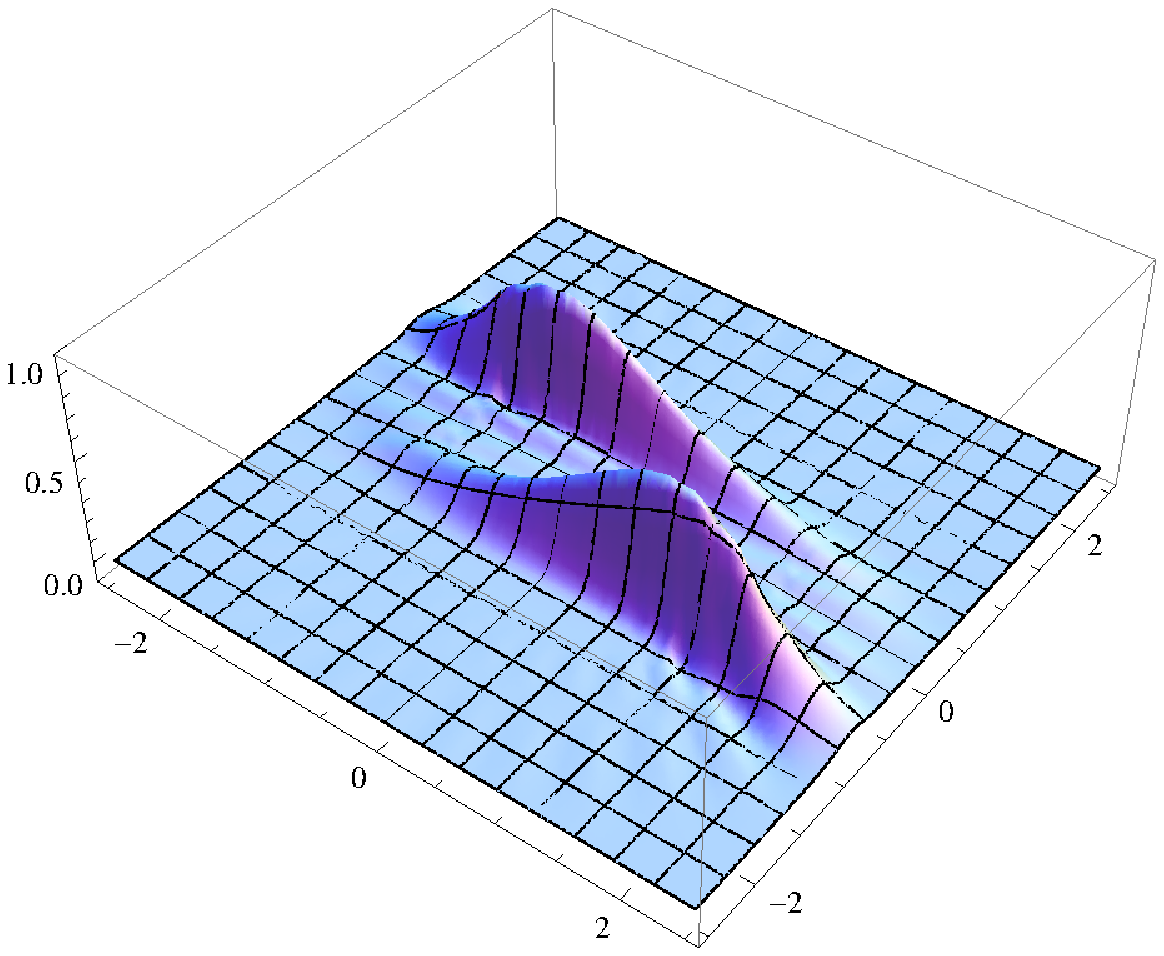,height=70mm,width=80mm}
\epsfig{figure=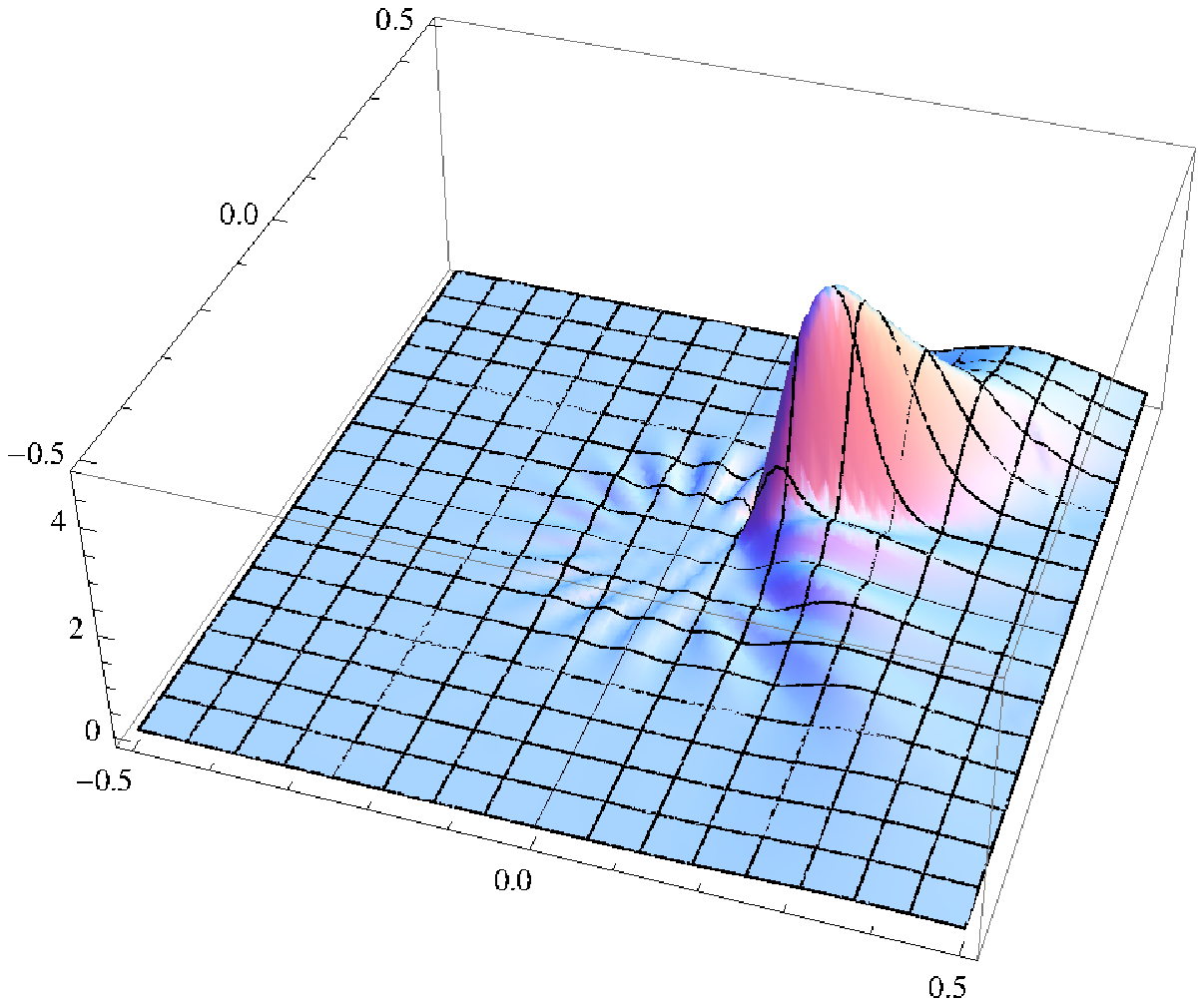,height=70mm,width=80mm}
\end{center}
\caption{The functions $\rho(x,y)$ (left) and $\sigma(X,Y)$ (right) for
  $S=\beta x^4/4$ with $\beta=0.1+i0.25 $ and $\lambda=0$. The matching is
  taken at $r_c=1.414$.  A $64$ point mesh is used for the interval $0<r<r_c$
  and $128$ for $0<R<R_c$. $32$ Fourier modes are used for $\arg Z$
  (corresponding to $64$ modes for $\arg z$). Numerically we find
$\sigma(0)= 0.011$ for this action.}
\label{fig:100}
\end{figure}
\begin{figure}[ht]
\begin{center}
\epsfig{figure=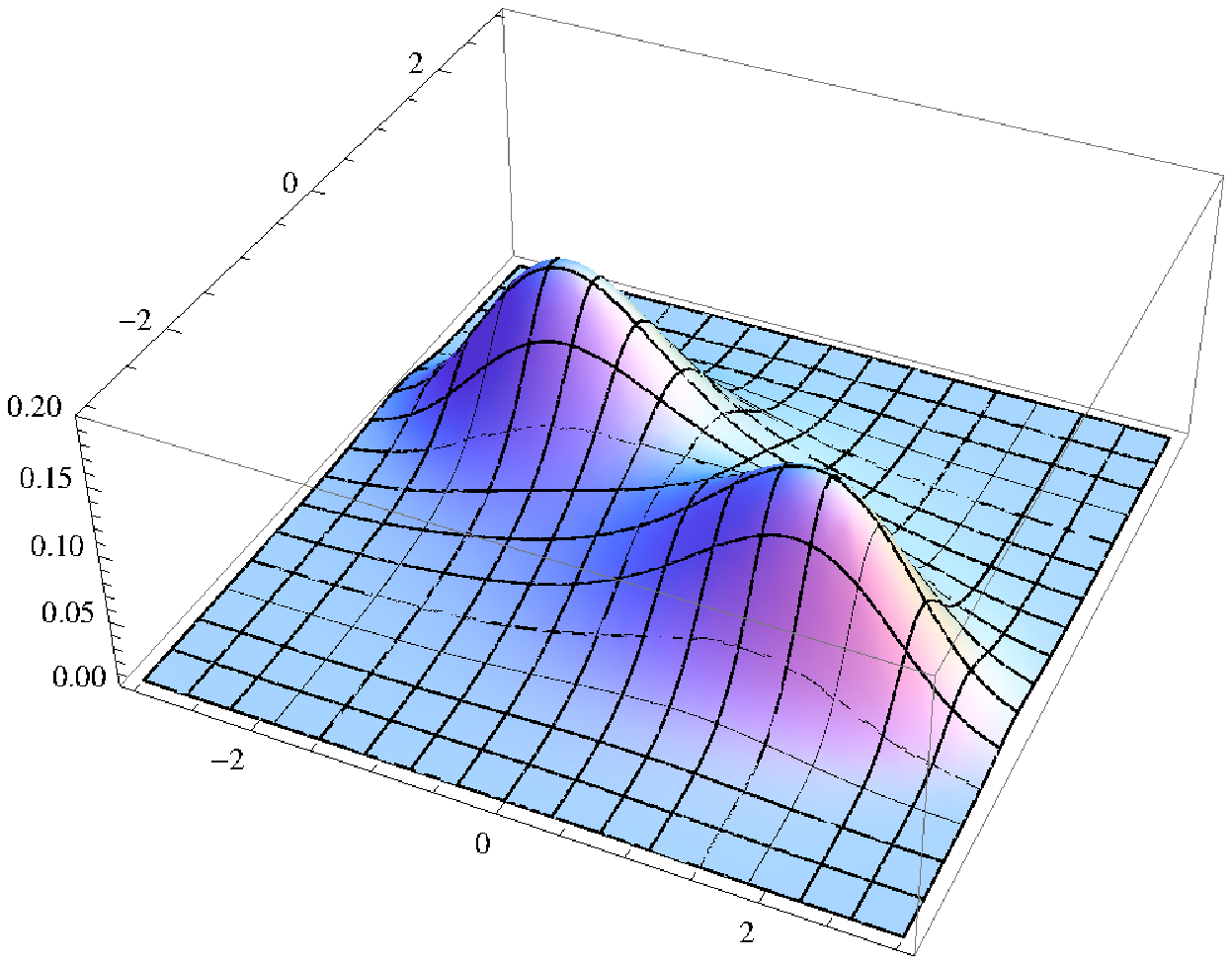,height=70mm,width=80mm}
\epsfig{figure=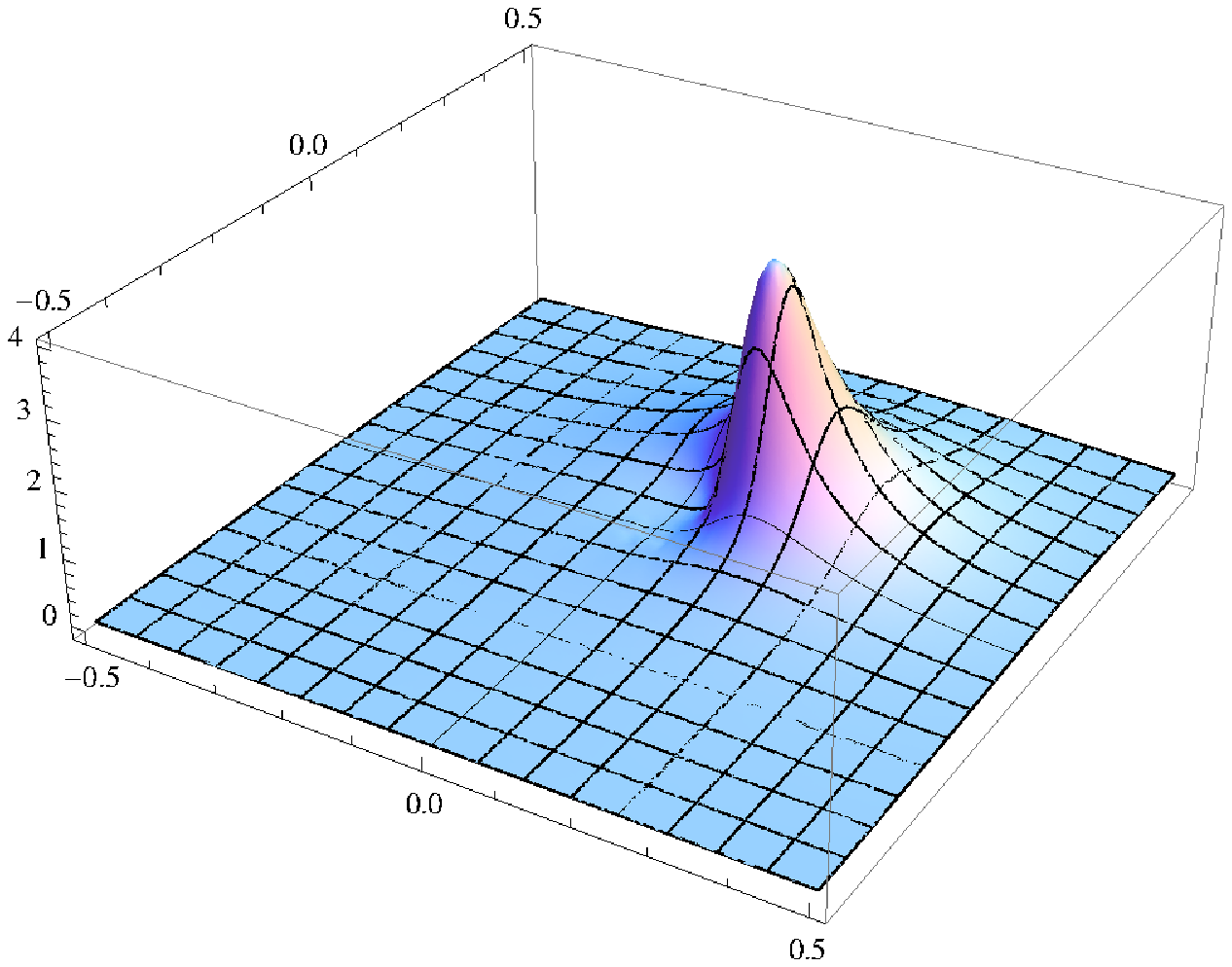,height=70mm,width=80mm}
\end{center}
\caption{Same as Fig. \ref{fig:100} for $\lambda=0.5$. Numerically
  $\sigma(0)=0.292$}
\label{fig:110}
\end{figure}
Using the two sets of coordinates, $z$ and $Z$, we have obtained numerical
solutions for various values of $\beta$ and $\lambda$. Similarly to the
treatment in Sec. \ref{sec:3.E}, the original coordinates, $(x,y)$, are used
on a disk $r=|z| \le r_c$ and the new coordinates $(X,Y)$ are used on
$R=|Z|\le R_c$ with $R_c=1/r_c^2$. The two solutions are matched at the
boundary, imposing the condition of zero net flux there. This condition fixes
the regularity condition at $Z=0$. Grids are used for $r$ and $R$ and a finite
number of Fourier modes are used for the angular variable. Numerical results
for $\rho(z)$ and $\sigma(Z)$ with $\beta=0.1+i0.5$ are displayed in
Fig. \ref{fig:100} for $\lambda=0$ and Fig.~\ref{fig:110} for
$\lambda=0.5$. As it would be expected, $\rho$ is smoother and wider for $\lambda=0.5$
and also $\sigma$ is larger near to $Z=0$.

Numerically we find an anomaly in the expectation values. The effect is more
visible for non vanishing $\lambda$ and this is the case we study in what
follows. In order to elucidate the presence or not of an anomaly in the
projected Fokker-Planck equation \Eq{3.13a}, we will analyze the behavior of
$\sigma(Z)$ in the region $Z=0$, equivalent to large $z$. From inspection of
the Fokker-Planck equation for $\sigma$ it follows that the diffusion does not
play a dominant role near $R=0$. In that region, the equilibrium equation can
be simplified to
\begin{equation}
0 \approx
 \beta \partial\sigma +  \beta^*\partial^*\sigma
=\beta_R \partial_X \sigma  + \beta_I \partial_Y \sigma
,
\qquad
\beta=\beta_R +i \beta_I
.
\label{eq:3.101}
\end{equation}
Defining rotated variables
\begin{equation}
u = \frac{X}{\beta_R} - \frac{Y}{\beta_I},
\quad
u^\prime = \frac{X}{\beta_I} + \frac{Y}{\beta_R}
,
\end{equation}
\begin{figure}[ht]
\begin{center}
\epsfig{figure=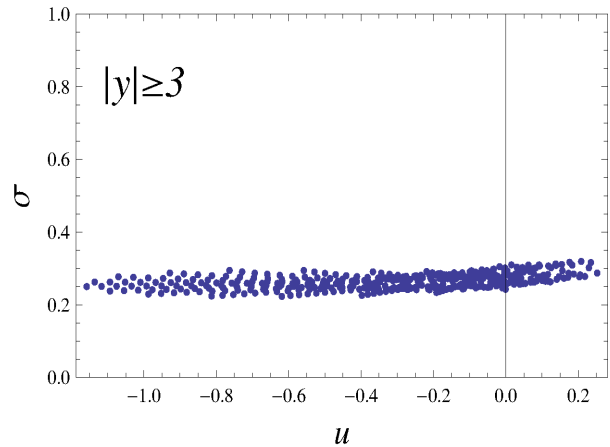,height=60mm,width=80mm}
\end{center}
\caption{For $\beta=0.1+i0.25$ and $\lambda=0.5$, values of the density
  $\sigma(Z)$ plotted against the scaling variable $u=X/\beta_R -
  Y/\beta_I$. The 444 values shown correspond to a subset of the points used
  in the numerical solution of the differential equation (except that the $32$
  Fourier modes have been traded for $32$ angular directions) defined by the
  condition $|y| \ge 3$. For these parameters $\sigma(0)=0.292$.}
\label{fig:120}
\end{figure}
\Eq{3.101} expresses that $\sigma$ depends only on $u$ and not on $u^\prime$.
We have verified this scaling in our numerical solution. In Fig.~\ref{fig:120}
we plot the values of $\sigma(X,Y)$ against $u$ in the region $|y|\ge 3$ for
the same action and $\lambda$ as in Fig.~\ref{fig:110}. The result shows
scaling, in the sense that $u$ alone determines the value of $\sigma$.

Fig.~\ref{fig:120} not only shows scaling in $u$ in the large $y$ limit, but
it also suggests that $\sigma(u)$ is actually rather flat. This can be
understood as a consequence of the diffusion: the flux is constant along the
flux lines and the diffusion (which is active not too close to $Z=0$) tends to
equate the flux on the different lines, making the flux to be nearly constant
everywhere. Near the origin, a constant flux implies a constant density
$\sigma$, since there the velocity is almost constant. This observation
suggests a simple model for the behavior of $\rho(x,y)$ in the region of large
$|y|$, namely (using \Eq{3.76})
\begin{equation}
\sigma \asymp \sigma(0)
,\qquad
\rho \asymp \rho_s(z) \equiv \frac{4\sigma(0)}{|z|^6}
\quad
(|y| \to \infty)
.
\label{eq:3.80}
\end{equation}

\begin{figure}[ht]
\begin{center}
\epsfig{figure=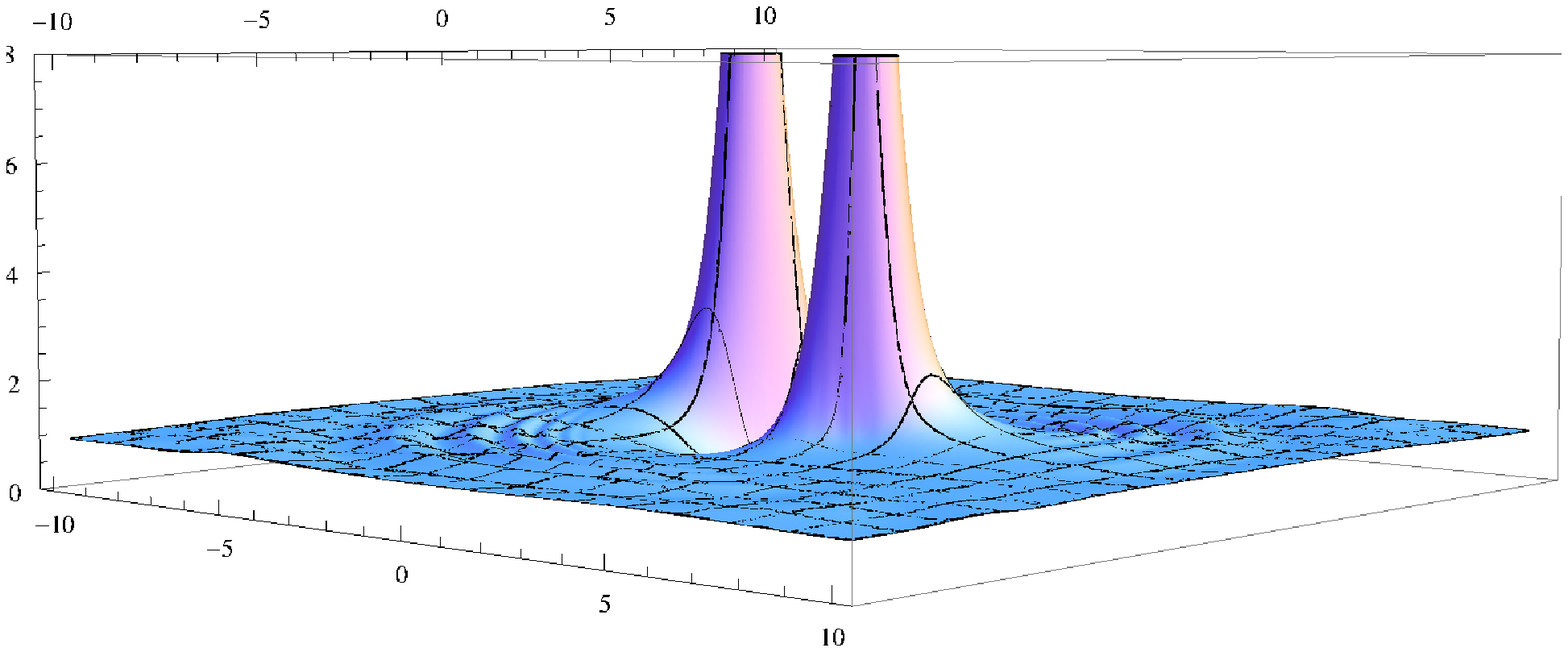,height=70mm,width=80mm}
\epsfig{figure=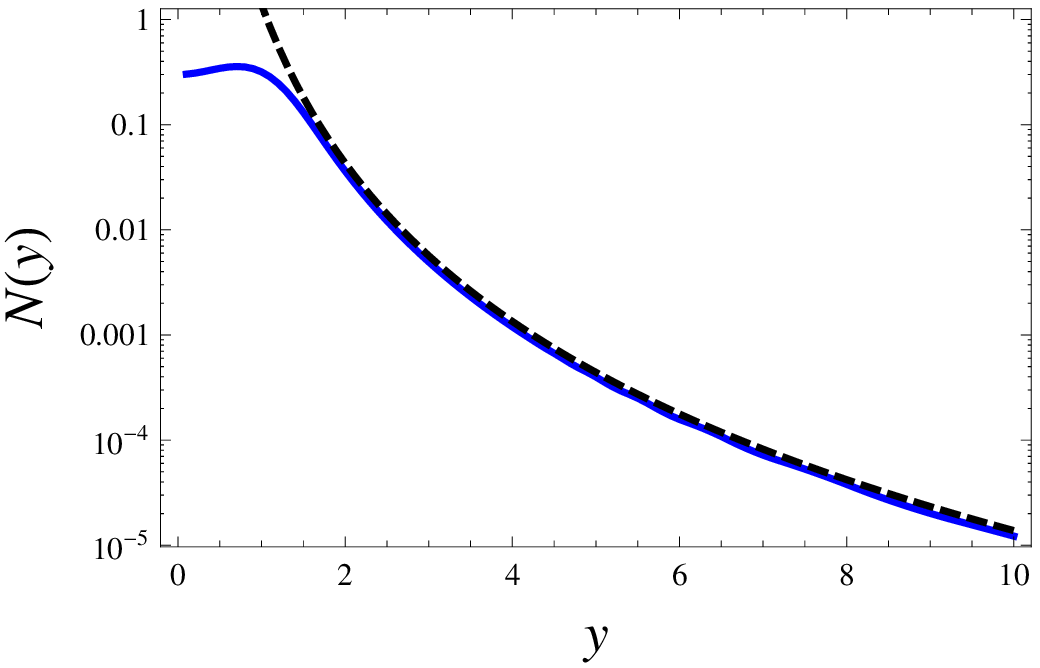,height=70mm,width=80mm}
\end{center}
\caption{For the solution of Fig.~\ref{fig:110}: $(a)$ Ratio
  $\rho/\rho_s=\sigma/\sigma(0)$ on the $z$ plane. $\rho$ is close to $\rho_s$
  in the asymptotic region. $(b)$ Marginal probabilities of $\rho$ (solid line,
  blue) and $\rho_s$ (dashed line, black). They only differ in the region
  $|y|<2$.}
\label{fig:130}
\end{figure}
The numerical validity of such asymptotic dependence is verified in
Fig.~\ref{fig:130}. Fig.~\ref{fig:130}$a$ represents the ratio $\rho/\rho_s$
on the $(x,y)$ plane. This ratio is close to unity outside a bounded region
around the origin $z=0$. Fig.~\ref{fig:130}$b$ shows the marginal density
$N(y) = \int dx \rho(x,y)$ compared to that corresponding to $\rho_s$, $N_s(y)
=3\pi\sigma(0)/(2y^5) $.

The combination of mathematical arguments and numerical evidence on the
validity of the asymptotic form in \Eq{3.80}, suggests that $\rho_s(z)$
contains the relevant information regarding the existence of an anomaly in the
projected Fokker-Planck equation. (Note however that we would really need to
estimate $\rho(x-iy,y)$ rather than $\rho(x,y)$.) In presence of the new
diffusion parameter $\lambda$, the anomaly (\ref{eq:2.17}) picks up an
additional term and it is changed to
\begin{equation}
\A(x) = -(v_y\rho+2i\lambda
\partial_z\rho)(x-iy,y)\big|_{y=-\infty}^{y=+\infty}
\equiv
\A(x)_+ - \A(x)_-
.
\label{eq:7.11}
\end{equation}
The new term with $\lambda$ is more convergent that the standard one
and so it is irrelevant to the anomaly. The parameter $\lambda$ has only an
indirect effect on the anomaly through its influence on $\rho(z)$. A direct
calculation assuming that $\rho_s$ saturates the anomaly gives
\begin{equation}
\A(x)_\pm =  \frac{2i\beta^* \sigma(0)}{x^3}
.
\label{eq:7.12}
\end{equation}
As expected the term with $\lambda$ has a vanishing contribution in the
$|y|\to\infty$ limit.

As it stands $\A(x)_+ = \A(x)_-$, and the anomaly would cancel. In fact,
$P(x)$ is an even function, and $\A(x)$ must also be even since parity is not
broken by the Fokker-Planck equation. Hence the $\A(x)_\pm$ just found, which
are odd functions, could never have a net contribution to the
anomaly. However, the result is singular at $x=0$ and a local distribution of
the same dimension, $\delta^{\prime\prime}(x)$, can be present upon
regularization. Such local distribution is an even function and would
contribute to the anomaly. In order to have a regularized result we turn to
the computation in momentum space. The anomaly in \Eq{7.11} takes the form
\begin{equation}
\tA(k)_\pm = 
-e^{ky}\left( \tilde{v}_y*\tilde{\rho}_s + \lambda (\partial_y - k) \tilde{\rho}_s
\right)
\Big|_{y=\pm\infty}
.
\end{equation}
Use of the relations $\tilde{z}* = i(\partial_k+y)$, \ $\tilde{z^*}* = i(\partial_k-y)$ in
$v_y= -(\beta z^3-\beta^* z^*{}^3)/2i$ and
\begin{equation}
\tilde{\rho_s}(k,y)
=4\pi\sigma(0) (k^2y^2+3|ky|+3) \frac{e^{-|ky|}}{8|y|^5}
,
\end{equation}
yields well defined distributions in momentum space, namely,
\begin{equation}
\tA(k)_\pm = 
\mp 2\pi \beta^* \sigma(0) k^2 \theta(\pm k)
,
\end{equation}
which correspond to
\begin{equation}
\A(x)_\pm =
2 i \beta^* \sigma(0) P\frac{1}{x^3}
\pm \pi \beta^* \sigma(0) \delta^{\prime\prime}(x)
.
\end{equation}
So finally, the assumption that $\rho_s$ saturates the anomaly yields
\begin{equation}
\A(x) =
2 \pi \beta^* \sigma(0) \delta^{\prime\prime}(x)
.
\label{eq:3.89}
\end{equation}
When this expression is introduced in the anomalous projected Fokker-Planck
equation at equilibrium,
\begin{equation}
0 = \partial_x (v P -\partial_x P) - a \delta^{\prime\prime}(x)
,\qquad
a \equiv 2 \pi \beta^* \sigma(0)
,
\end{equation}
the normalized biased solution is obtained as
\begin{equation}
P(x) = (1+a) P_0(x)  - a \delta(x)
\end{equation}
where $P_0(x)$ is the normalized unbiased solution $P_0\propto e^{-\beta
  x^4/4}$.

For a generic observable $\cO(x)$, the anomalous expectation value becomes
\begin{equation}
\esp{\cO} = (1+a)\esp{\cO}_0 - a \cO(0)
,
\end{equation}
where $\esp{\cO}_0$ is the unbiased result. In particular\footnote{
$\esp{x^n}_0 = (4/\beta)^{n/4} 
\Gamma(\frac{n+1}{4})/\Gamma(\frac{1}{4})$ for even nonnegative $n$.}
\begin{equation}
\esp{x^n} = (1+a)\esp{x^n}_0, \qquad n>0
.
\label{eq:7.23}
\end{equation}
This relation can be tested numerically. For $\beta=0.1+i0.25$ and
$\lambda=0.5$, ~ $a= 0.184 - 0.459 i$, and we obtain
\ignore{
\begin{equation}\begin{split}
\esp{x^2}_0 &= 1.079 - 0.730 i,
\quad
\esp{x^2} = 0.946 - 1.238 i,
\quad
(1+a)\esp{x^2}_0 = 0.942 - 1.359 i
, 
\\
\esp{x^4}_0 &= 1.379 - 3.448 i ,
\quad
 \esp{x^4} = 0.107 - 4.664 i,
\quad
(1+a)\esp{x^4}_0 = 0.050 - 4.714 i
.
\end{split}\end{equation}
}
\begin{align}
\esp{x^2}_0 &= 1.079 - 0.730 i,
&
\esp{x^4}_0 &= 1.379 - 3.448 i ,
\nonumber
\\
\esp{x^2} &= 0.946 - 1.238 i,
&
\esp{x^4} &= 0.107 - 4.664 i,
\label{eq:7.24}\\
(1+a)\esp{x^2}_0 &= 0.942 - 1.359 i,
&
(1+a)\esp{x^4}_0 &= 0.050 - 4.714 i
\nonumber
\end{align}
These results show that there is indeed a bias in complex Langevin
solution. Furthermore, the anomaly estimated from \Eq{3.89} gives a fair
description of the bias in $\esp{x^2}$ and $\esp{x^4}$. For higher powers
the numerical errors accumulate and it is harder to draw definite conclusions.
Also, in the most interesting case of $\lambda=0$, $\sigma(0)$ is too small
and any possible bias competes with the numerical error of the calculation.

It is possible that asymptotic subleading corrections to $\rho_s$ could give a
contribution to the anomaly. An easy calculation generalizes the result in 
\Eq{3.89}. Specifically, a subleading term of the type
\begin{equation}
\frac{c_{n,m}}{|z|^6}\frac{1}{z^n z^*{}^m}
\end{equation}
in $\rho$ would yield a contribution to $\A(x)$ equal to
\begin{equation}
c_{n,m}\pi\beta^* \frac{(-1)^n}{(n+2)!} \delta_{m,0} \delta^{(n+2)}(x)
.
\end{equation}
Such terms would modify the expectation values of higher powers of $x$.

Note that we have not directly applied the analytic projection operator $K$ to
$\rho_s(z)$ to obtain an estimate of $P(x)$. From dimensional considerations,
the projection of $\rho_s(z)$ would yield a $\delta^{(4)}(x)$ term. However,
when this is analyzed in momentum space, it can be seen that the $k^4$ term
comes from the small $y$ region, where the estimate $\rho_s(z)$ is not
reliable.  If the small $y$ region is removed one obtains instead $k^2$, i.e.,
a $\delta^{\prime\prime}(x)$ estimate for $P(x)$. The problem with this
procedure is that it is not clear which part of the estimate is anomalous
and which part is just a regular contribution from $P_0(x)=e^{-S(x)}$.

\section{Summary and conclusions}
\label{sec:con}

As noted in the Introduction, very general complex probability distributions
admit a representation on the complexified manifold.  The goal of the complex
Langevin method is to construct one such valid representation for
$e^{-S(x)}$. While the method is certainly handy and elegant, in the
Introduction it is shown that perfectly valid complex actions cannot be
reproduced with this approach, just from an analysis of the support of
$\rho_{\rm CL}(z)$ on the complexified manifold. Also it is shown that barring
the cases of quadratic or real actions, the only known exact solution of a
stationary Fokker-Planck solution [see Eqs. (\ref{eq:3.5}) and (\ref{eq:3.6})]
gives completely incorrect results, even for the two Fourier modes for which
the integrals are absolutely convergent.

It was already known \cite{Salcedo:1993tj} that even if the projected
Fokker-Planck equation is the naive one, spurious solutions of it can be
selected by the algorithm. This is briefly reviewed in
Sec. \ref{sec:spu}. However, in practice such spurious solutions have to be
enforced by using a suitable kernel or by choosing a complex probability with
zeros or poles on the complex plane. A more pressing problem is discussed in
Sec. \ref{sec:anom}, namely, the emergence of an anomaly in the projected
Fokker-Planck equation [see Eqs. (\ref{eq:3.13a}) and (\ref{eq:2.17})]. The
anomaly $\A(x)$ is a surface term which would be absent if the
stationary density $\rho(z)$ were sufficiently convergent far from the real
manifold. Similar boundary terms have been described in the analysis of
\cite{Aarts:2009uq}. Here we provide a precise mathematical form to the
anomaly which allows us to derive explicit anomalous relations between the
behavior at infinity and the bias induced on the observables
[see e.g., Eqs. (\ref{eq:3.43}), ~(\ref{eq:6.8}), or~(\ref{eq:7.23})].

In Sec. \ref{sec:anu1} we analyze the one-dimensional periodic action
$\beta\cos(x)+imx$ (with real $\beta$ and integer $m$). This action is of
interest because in a Monte Carlo simulation using complex Langevin, correct expectation values are
obtained for $\cos(x)$ and $\sin(x)$.  Also, the flow on the complex plane
looks healthy and with suitable located fixed points (see
Fig. \ref{fig:5}$b$), yet this is not guarantee of a correct behavior. As
noted, the flow would be qualitatively similar when $m$ is not an integer, and
in that case the stochastic process would produce some necessarily periodic
distribution which certainly would not be a representation of
$e^{-\beta\cos(x)-imx}$, which is not a periodic function. Also, the anomaly
depends on how often the Langevin walkers visit the remote regions far from
the real manifold and this is not obvious from the topology of the flow.  As
it turns out, we are able to show that the anomaly does not affect the
expectation values of Fourier modes with $k=-1,0,1,2,3,\ldots$ (for positive
$m$), and hence the complex Langevin method provides a stationary solution
which correctly reproduces all those modes. However, an anomaly is present and
all the $k<-1$ modes are biased. This warns us that conclusions based solely
on numerical experiments could be misleading.

That an anomaly is necessarily present for the action $\beta\cos(x)+imx$ is
proven in detail in Sec. \ref{sec:proof}. For positive $m$, the support of
$\rho(z)$ (the stationary solution of the complex Langevin process) lies
entirely on the lower half plane. Our approach is to make a change of
variables $(x,y)$ to $(X,Y)$ which effectively compactifies the large negative
$y$ region to a point; this is the origin in the new variables. We show that
this point is perfectly regular for the stochastic process in the new
variables [see \Eq{3.48} and Fig. \ref{fig:7}], and in particular the density
$\sigma(X,Y)$ (related to $\rho(x,y)$ by the Jacobian of the change of
variables) is strictly non zero at $(X,Y)=0$. This suffices to show that there
is an anomaly in the projected Fokker-Planck equation for this action [see
  \Eq{3.44}]. Further, we numerically solve the differential equation using a
finite number of Fourier modes for the periodic variable $x$ and a mesh for
$R$ (related to $y$). This allows us to verify numerically all the relations
previously derived.

The study of anomalies in more general periodic actions is addressed in
Sec. \ref{sec:3.E}. It is argued that the anomaly does not vanish in the
general case. We analyze in some detail the actions of the form $\beta
\cos(x)$ for complex $\beta$. Once again, we use compactifying coordinates
$(X,Y)$ and show that the anomaly is controlled by the behavior of
$\sigma(X,Y)$ near $X=Y=0$ [see \Eq{3.66}]. For these actions the support of
the stationary solution of the the complex Langevin process is no longer
restricted to a half-plane [although parity is preserved by $\rho(z)$]. So we
use simultaneously the original coordinates $(x,y)$ and the new coordinates
$(X,Y)$, in two juxtaposed patches, in order to solve numerically the
Fokker-Planck equation. The two partial solutions are matched by requiring
that there is no net flux through the boundary of the patches. The numerical
solution allows to establish the unequivocal presence of an anomaly, and
moreover, the corresponding predicted bias is correctly reproduced, at a
numerical level [see \Eq{6.9}].

Finally, in Sec. \ref{sec:nonp}, a non periodic action is analyzed, namely,
$S(x) = -\beta x^4/4$, for $\Re\beta>0$. In this case, in order to compactify
and regularize the problem at infinity, the change of variables $Z=1/z^2$ is
applied. These coordinates cover half of the $z$-plane but this is sufficient
since $\rho(z)$ is an even function. In this way the Fokker-Planck equation is
solved numerically in two patches (one including $z=0$, the other $Z=0$) and
the solutions are matched at the common boundary.  Once again the behavior of
$\sigma(Z)$ near $Z=0$ is expected to determine the presence of an
anomaly in the projected equation, however, the value found for $\sigma(0)$ is
numerically small, and this prevents us to establish in an unambiguous way
whether an anomaly is present or not. The same situation takes place in the
expectation values: any possible bias competes with the numerical error in the
calculation. In order to obtain a clearer case, an extra diffusion is added in
the complex Langevin process controlled by a positive parameter $\lambda$ [see
  \Eq{2.2a}]. Such term has been considered before in similar studies
\cite{Aarts:2009uq,Aarts:2011ax}. The extra diffusion, although formally
correct, has a negative effect on the complex Langevin process as a Monte
Carlo method, yet simultaneously it greatly improves the mathematical behavior
of the stationary solution, regarding its analyticity. This is advantageous both
for the numerical solution of the differential equation and for the analysis
of asymptotic behaviors. As a consequence, for positive $\lambda$, we are able
to unambiguously establish the existence of an anomaly and a bias in the
expectation values [see \Eq{7.24}]. In fact, from the asymptotic analysis we
obtain an analytical form for the anomaly [see \Eq{3.89}] which is fully
confirmed by the numerical calculation.\footnote{Actually, we first observed
  empirically the relation in \Eq{7.23}, with $a=c\beta^* \sigma(0)$ and
  $c\approx 6$.}

%____________________________
One could wonder if a better estimate would be obtained for the observables by
chopping off somehow the anomaly-generating contributions in the $\rho(z)$
produced by complex Langevin. It is difficult to give a complete answer
without entering into casuistic. Nevertheless, there are two cases which are
quite clear. For the action $S = x^4/8+2ix$ noted in the Introduction, after
stabilization the support of $\rho(z)$ lies entirely below the real
axis. Removing part of the support does not change this. So the expectation
value of $e^{-ix}$ would still be smaller that unity (in absolute value),
while the correct result is roughly $-5$. Therefore the error would be sizable
anyway. Another clear case is that of $S = i \cos(x)$, with exact solution for
the Fokker-Planck equation $\rho(x,y) \propto \mathrm{sech}^2(y)$. This
(incorrectly) predicts the vanishing of all expectation values of $e^{-ikx}$
(for $k$ different from $0$).  If we remove all points in the complex plane
with $|y|$ larger than some cutoff, still the new $\rho$ will be a function of
$y$ only, and it will produce the same incorrect expectation values.
%______________________________

Since the anomaly is due to too frequent visits of the Langevin walker to
remote regions far from the real manifold, the use of new variables, in which
$y=\infty$ is a regular point, has played an important role our in the
analysis of the anomaly.  The choice of the new variables is such that for
them $y=\infty$ becomes a finite point, say $Z=0$, but more importantly, the
velocity of the drift in the new variables, $V(Z)$, is strictly finite in a
neighborhood of $Z=0$, that is, neither zero nor infinite. Also, the diffusion
becomes negligible there, because a finite stochastic jump in $z$ is a small
jump in $Z$. This can be done systematically by choosing $Z(z)$ such that (we
discuss the one-dimensional case only, however, similar changes of variables
can be carried out in higher dimensions)
\begin{equation}
\frac{dZ}{dz} = \frac{1}{v(z)}
,
\label{eq:8.1}
\end{equation}
which yields ($Z=X+iY$)
\begin{equation}
-\partial_t \sigma = - \frac{1}{J}\partial_x^2(J\sigma) + \partial_X\sigma
,
\end{equation}
with
\begin{equation}
\rho = J \sigma
,\qquad
J= \frac{1}{|v(z)|^2}
.
\end{equation}
Up to corrections from the diffusion, this implies that the drift velocity $V$
is constantly equal to $1$, so the walkers move uniformly to the right.
Essentially, these are the variables used in \Eq{7.5} for $S=\beta x^4/4$, and
also in \Eq{3.41} for the periodic actions.\footnote{Actually, a velocity
  $v=\sin(z)$ yields $Z=\log(i\tan(z/2))$ ~(with $\log(1)=0$ and $-\pi < \arg
  < \pi$), whereas the change in \Eq{3.41}, is $W = 2e^{ix+y}/(1-e^{2y})$,
  however $Z=\log((W + 1 -\sqrt{|W|^2+1})/(W - 1 +\sqrt{|W|^2+1})) = -W^* +
  \frac{1}{12}(3W -W^*)W^{ * 2} + O(|W|^5 ) $ so the change of variables
  between the variables in \Eq{3.41} and \Eq{8.1} is
  regular and real-analytic at $Z=0$.  }

\begin{figure}[ht]
\begin{center}
\epsfig{figure=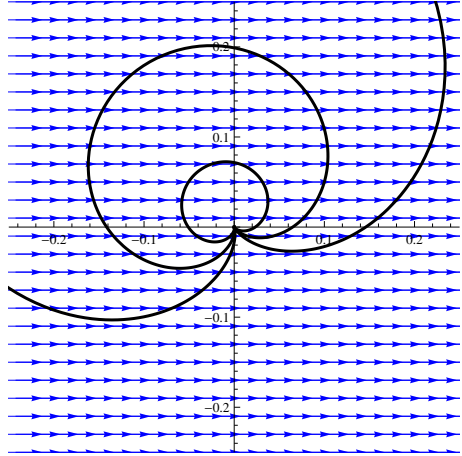,height=60mm,width=60mm}
\end{center}
\caption{Velocity field of $S = \beta x^4/4$ in the variable $Z=1/(2\beta
  z^2)$, for $\beta=0.1+i0.25$. The solid curves correspond to lines of
  $x=$constant, for $x=2$, $3$, and $5$. The diffusion randomly moves the
  Langevin walkers along these lines allowing them to visit a neighborhood of
  $Z=0$. This suggests that $\sigma(0)$, or at least its limit from negative
  $X$ if this function is not continuous at $Z=0$, will not be exactly zero.
(The peak of $\sigma(X,Y)$ displayed in Fig.~\ref{fig:100} falls at $Z=0.32-i0.18$ here.)}
\label{fig:10}
\end{figure}
When the Langevin walker follows an orbit of $v(z)$ (let us obviate the
diffusion at the moment) and this orbit visits a very remote region on the
complex plane $z$ before returning, in the $Z$-plane this corresponds to a
regular orbit passing near $Z=0$. The fact that the velocity $V$ is strictly
finite there suggests that $\sigma(0)$ will not be infinite (which would
follow from $V=0$) nor zero (which would follow from $V=\infty$). The point
$Z=0$ (i.e., $z=\infty$) is a regular point and there is an orbit passing
through it taking a finite Langevin time to reach $\infty$ and come back. The
diffusion introduces irregularities but its effect is small in a neighborhood
of $Z=0$.  Actually, the diffusion should help to flatten $\sigma$ there: in
general the walkers will have a chance to be moved to $X<0$ by diffusion (this
is confirmed by studying the lines $x=$constant on the $Z$-plane in the
previous examples, see Fig.~\ref{fig:7} and Fig.~\ref{fig:10}) and so
inevitably the constant drift will move them to a neighborhood of $Z=0$,
implying that $\sigma$ should be finite there. All this indicates that
$\sigma$ might be continuous and non vanishing at $Z=0$ or at least have a
finite limit from $X<0$. The value of $\sigma(0)$ might be exceedingly small
in practice but nevertheless it would signal a bias in the algorithm.

Because we have seen in the various cases analyzed above that a non null
$\sigma(0)$ is tied to an anomaly, the arguments just presented tend to
suggest that a non null anomaly, and so a bias in the estimates of the
expectation values of the observables, would be the rule rather than the
exception in the complex Langevin dynamics. An obvious objection is that the
change of variable breaks down at the fixed points of $v$, however, this is
not a problem if the set of fixed points is bounded: in this case the change
of variable is still well defined in a neighborhood of $Z=0$, and a different
patch can be used elsewhere. Alternatively, one can work on the Riemann
surface; this is technically hard, but it does not invalidate the
construction. A more substantial critique is that $y=\infty$ needs not
correspond to a finite $Z$.  That the infinity of $z$ can be brought to $Z=0$
follows from the explicit solution of \Eq{8.1}
\begin{equation}
Z = \int^z_{\infty} \frac{dz}{v}
\label{eq:8.4}
,
\end{equation}
for points $z$ lying beyond all the fixed points.\footnote{Here we are
  assuming that $y= \pm\infty$ have neighborhoods free from fixed
  points. These two patches need not overlap. A variable $Z$ is defined in each patch.}
A problem arises if the integral is not convergent. This is the case for a
quadratic action, since $v\sim z$ for large $z$, and the integral diverges
logarithmically. The integral converges provided $v$ increases faster than
$|z|$ as $|z|\to \infty$. There is an intuitive explanation for this behavior.
In fact, $Z$ in \Eq{8.4} represents the time that the walker needs to reach
infinity following an orbit of the field velocity. The integral converges if
this time is finite. It seems to be a sensible criterion that when it takes
only a finite time for the Langevin walker to arrive to infinity, the visits
there will be frequent and $\rho(z)$ will falloff much to slowly for the
integration by parts to be justified. This introduces an anomaly and a bias in
the expectation values. For a quadratic action the construction does not go
through and an anomaly does not arise.

In order to investigate whether the anomaly could be obtained in closed form,
let us momentarily assume that the regions $y = \pm\infty$ are sufficiently
well described by $\sigma(Z)\approx \sigma_\pm(0)$, where we allow two
different values in the two regions. In this case the density can be
approximated as
\begin{equation}
\rho(z) \asymp \rho_{s,\pm}(z) \equiv \frac{\sigma_\pm(0)}{|v(z)|^2}
,\qquad
y \to \pm \infty
.
\end{equation}
This formally yields for the anomaly, using \Eq{2.17},\footnote{The result in \Eq{7.12} follows
  this scheme, albeit with a different normalization in $\sigma$ since there
  $Z=1/z^2$ instead of $1/(2\beta z^2)$.}
\begin{equation}
\A(x)_\pm = 
-\left(
\frac{v-v^*}{2i}\frac{\sigma_\pm(0)}{vv^*} 
\right)(x-iy,y)
\Big|_{y=\pm\infty}
=
-\frac{\sigma_\pm(0)}{2i}\left(
\frac{1}{v(x+2iy)^*} - \frac{1}{v(x)}
\right)
\Big|_{y=\pm\infty}
=
\frac{1}{2i}\frac{\sigma_\pm(0)}{v(x)}
.
\label{eq:8.6}
\end{equation}

This calculation would indicate that there is an anomaly if and only if
$\sigma_+(0) \not= \sigma_-(0)$. [An exception would perhaps occur at the
  points where $v(x)=0$, as seen in Sec. \ref{sec:nonp}.] In particular, for
even actions there would be no anomaly. However, we have shown that
$S=\beta\cos(x)$, for instance, is anomalous. The paradoxical result is
avoided by noting that the asymptotic relation $\rho \asymp \rho_s$ refers to
the limit $y\to \infty$ with $x$ fixed, while the anomaly requires to have a
control on $\rho$ in the limit $\rho(x-iy,y)$ as $y\to \infty$ with $x$ fixed.
This can be seen for the action $S=i\beta_I \cos(x)$ for which $\rho$ is known
in closed form. In this case $\sigma_\pm(0)=\beta_I^2/4\pi$, so
\begin{equation}
\rho(x,y) = \frac{1}{4\pi\cosh^2(y)} \asymp \frac{e^{-2|y|}}{\pi}
,\qquad
\rho_s(x,y) = \frac{\beta_I^2/4\pi}{\beta_I^2 |\sin(x+iy)|^2}
 \asymp \frac{e^{-2|y|}}{\pi}
,
\end{equation}
and $\rho_s$ correctly accounts for the behavior of $\rho$ for large $y$. On
the other hand, with fixed $x$ and $y\to \pm \infty$,
\begin{equation}
\rho(x-iy,y) \asymp \frac{e^{-2|y|}}{\pi}
,\qquad
\rho_s(x-iy,y) \asymp \mp \frac{e^{\mp i x}}{2i\sin(x)}
\frac{e^{-2|y|}}{\pi}
.
\end{equation}
The incorrect extra factor in $\rho_s(x-iy,y)$, as compared to $\rho(x-iy,y)$,
produces the incorrect expression in \Eq{8.6} instead of \Eq{3.21}.

The previous discussion indicates that a reasoning that correctly estimates
$\rho(x,y)$ in the $y = \pm\infty$ regions does not necessarily suffice for
describing the behavior of $\rho(x-iy,y)$, as required in the anomaly
expressions. Nevertheless, it remains true that, in all cases analyzed, a non
vanishing value of $\sigma(0)$ is tied to the presence of an anomaly and a
bias in complex Langevin results, and we have argued that this will be the
case quite generally. The numerical results do not allow to clearly establish
whether $\sigma(0)$ vanishes or not for $S=\beta x^4/4$ with $\lambda=0$ (the
interest of $\lambda>0$ is rather academic) but it seems clear that a positive
value of $\sigma(0)$ would yield an anomaly also in this case.

Another question is how these findings are modified by increasing the number
of variables. Here we have studied one-dimensional problems, concluding that
an anomaly is present quite generally, and this conclusion immediately extends
to ultralocal actions, however, it is conceivable that in general the strength
of the anomaly might depend on the dimension of the configuration manifold. In
the instances in which the anomaly were decreasing to zero in the continuum or
thermodynamic limits, the bias would eventually be screened by the standard
Monte Carlo fluctuations and the method would be valid there.  Also, a kernel
modifies the algorithm while keeping its formal validity. It would be
interesting to explore whether a suitable kernel could keep the walkers close
to the real manifold thereby quenching or removing the anomaly.  These points
deserve further study.

%________________________________________
We want to briefly comment on the relation between this work and the method of
Lefschetz thimbles \cite{Cristoforetti:2012su}. Like complex Langevin, this
technique to attack the sign problem also relies on the fact that the action
is a holomorphic function. It amounts to trade the original real submanifold
$\R^n$ (embedded within the complex manifold of complex configurations $\C^n$)
by one or more cleverly chosen submanifolds of the same dimension, the
Lefschetz thimbles, whose combination is homologous to the real submanifold,
in the sense that holomorphic functions weighted with the Boltzmann factor of
the action have the same integral in both treatments.  (A smeared version of
this technique is studied in \cite{Fukushima:2015qza}, another interesting
extension can be found in \cite{Alexandru:2015sua}.)

In the complex Langevin deterministic flow, non degenerated fixed points are
attractive or repulsive (or neutral) but not saddle points. Lefschetz thimbles
are based on the gradient flow of $\Re(S)$ which shares the same fixed points,
but now as saddle points. Passing through each fixed point there are two
$\Im(S)={\text{constant}}$ submanifolds, one for which the fixed point is a
maximum of $\Re(S)$ and another for which is a minimum, the latter is the
(stable) Lefschetz thimble associated to the fixed point (the other
submanifold being the unstable thimble). This choice of integration manifold
generalizes the stationary phase approximation. In practice, a subset of the
(stable) thimbles have to be joined to form a manifold homologous to $\R^n$
(namely, those whose fixed point have the unstable thimble intersecting $\R^n$)
\cite{Witten:2010cx}.

Several studies have analyzed the relation between both techniques and how a
wrong convergence of complex Langevin could be understood from the Lefschetz
thimble point of view (see e.g. \cite{Aarts:2014nxa}). Although this is not a
fixed rule, the support of the complex Langevin distribution tends to follow
the Lefschetz thimble and give better results when there is a single dominant
thimble. (A technique to enforce this feature as a way to improve complex
Langevin is studied in \cite{Tsutsui:2015tua}.)  Ref. \cite{Hayata:2015lzj}
finds that complex Langevin fails if multiple Lefschetz thimbles dominantly
contribute with different complex phases. In fact, if there is just one
dominant thimble this will usually correspond to the basin of a stable fixed
point of the complex Langevin dynamics, however, if additional thimbles
corresponding to {\em repulsive} points are relevant, such regions will not be
visited by the Langevin walker and one can expect a bias in the estimates. For
another argument related to the present work,\footnote{I thank an anonymous
  referee for providing this argument.} the thimbles are boundaryless
non-compact $n$-dimensional manifolds in $\C^n$. Keeping those having a
contribution and with suitable orientation, their union forms a manifold
homologous to $\R^n$. Typically when there is more than one (contributing)
fixed point, the thimble stems from the fixed point towards infinity and there
joins with the thimble of another fixed point.\footnote{The union of
  thimbles is homologous to $\R^n$ so they have to join somehow, but each
  thimble carries a different value of $\Im(S)$, the only solution is their
  joining at infinity.} Although the deterministic flow of complex Langevin
and the gradient flow of $\Re(S)$ are different, they coincide asymptotically
\cite{Aarts:2014nxa}. This means that when two relevant thimbles join at
infinity they do so at a runaway trajectory of the Langevin walker. If the
walker has to visit both thimbles for a proper sampling it necessarily must
expend some time at infinity and this introduces an anomaly. This would be a
relation between the problems observed in complex Langevin when there is more
than one dominant thimble and the presence of an anomaly in the projected
Fokker-Planck equation. This heuristic argument implicitly assumes that a proper
sampling can only be achieved when the support of the complex Langevin walker
is close to the support of the relevant thimbles. A possible caveat though is
that each thimble carries a different global phase and also a local phase
through the Jacobian since the thimbles are not horizontal submanifolds in general. In other words, the
thimbles do not define a proper representation with positive weight as defined
in the Introduction, and this would account for the different supports in the
two treatments. In this regard, it is interesting to point out that in
\cite{Salcedo:2007ji} it is shown that any periodic action in any number of
dimensions, can be represented using just two horizontal submanifolds, with
positive weights on them, and the method has been extended to non periodic
actions  in
\cite{Salcedo:2015jxd} (albeit in the one-dimensional case). In this view, the
study of the systematic construction of such two-branch representation seems
worthwhile.

%____________________________________

\begin{acknowledgments}
This work was supported by Spanish Ministerio de Econom\'{\i}a y
Competitividad and European FEDER funds (grant FIS2014-59386-P), and by
the Agencia de Innovaci\'on y Desarrollo de Andaluc\'{\i}a (grant FQM225).
\end{acknowledgments}

%\appendix

\end{document}